\documentclass[11pt]{article}

\usepackage[square, numbers, comma, sort&compress]{natbib}

\usepackage{graphicx,amsmath,amssymb}
\usepackage{indentfirst}
\usepackage[usenames]{color}
\usepackage[colorlinks=true, urlcolor=navyblue, linkcolor=navyblue, citecolor=navyblue]{hyperref}
\usepackage{epstopdf}
\usepackage{enumerate}
\usepackage{appendix}
\usepackage{lineno}
\usepackage{setspace}
\usepackage{wrapfig}
\usepackage{lipsum}
\usepackage{placeins}
\usepackage{ulem}
\usepackage{caption}
\usepackage{appendix}
\definecolor{navyblue}{rgb}{0,0.08,0.45}

\usepackage{dcolumn}
\usepackage{bm}
\usepackage{xfrac}
\usepackage{dsfont}
\usepackage{lipsum}
\usepackage{braket}
\usepackage{times}
\usepackage{xcolor}

\makeatletter
\providecommand{\doi}[1]{%
  \begingroup
    \let\bibinfo\@secondoftwo
    \urlstyle{rm}%
    \href{http://dx.doi.org/#1}{%
      doi:\discretionary{}{}{}%
      \nolinkurl{#1}%
    }%
  \endgroup
}
\makeatother

\begin{document}

\begin{flushright}
{\small Preprint Nos.\ JLAB-PHY-24-4003;\; SLAC-PUB-17762;\; NJU-INP 088/24
}
\end{flushright}

\vspace{33pt}

\centerline{\large \bf {Poincar\'e invariance and the Unruh effect} }

\vspace{40pt}

\centerline{Alexandre~Deur$^*$}

\vspace{5pt}

\centerline {\it Thomas Jefferson National Accelerator Facility, Newport News, Virginia 23606, USA}

\vspace{15pt}

\centerline{Stanley J. Brodsky}

\vspace{5pt}

\centerline {\it SLAC National Accelerator Laboratory, Stanford University, Stanford, California 94309, USA}

\vspace{15pt}

\centerline{Craig D.~Roberts}

\vspace{5pt}

\centerline {\it School of Physics, Nanjing University, Nanjing, Jiangsu 210093, China}
\centerline {\it Institute for Nonperturbative Physics, Nanjing University, Nanjing, Jiangsu 210093, China}

\vspace{15pt}

\centerline{Bal{\v s}a Terzi{\'c}}

\vspace{5pt}

\centerline {\it Department of Physics, Old Dominion University,
Norfolk, Virginia 23529, USA}

{\small \centerline{\today}}

\vspace{15pt}

\vspace{60pt}

\abstract{
In quantum field theory, the vacuum is popularly considered to be a complex medium populated with virtual particle + antiparticle pairs.
To an observer experiencing uniform acceleration, it is generally held that these virtual particles become real, appearing as a gas at a temperature that grows with the acceleration.
This is the Unruh effect. 
However, it has been shown that vacuum complexity is an artifact, produced by treating quantum field theory in a manner that does not manifestly enforce causality.
Choosing a quantization approach that patently enforces causality, the quantum field theory vacuum is barren, bereft even of virtual particles.
We show that acceleration has no effect on a trivial vacuum; hence, there is no Unruh effect in such a treatment of quantum field theory. 
Since the standard calculations suggesting an Unruh effect are formally consistent, insofar as they have been completed, there must be a cancelling contribution that is omitted in the usual analyses.
We argue that it is the dynamical action of conventional Lorentz transformations on the structure of an Unruh detector.
}


\section{Introduction} 
The Unruh effect is the name given to a theoretical prediction that an accelerated observer can detect particles which are unobservable to an inertial observer \cite{Fulling:1972md, Davies:1974th, Unruh:1976db, Crispino:2007eb}.
The effect is often interpreted as a consequence of the intrinsic complexity of the inertial frame vacuum, which is commonly imagined to be populated by pairs of virtual particles that become detectable in an accelerated frame.
However, vacuum complexity can be viewed as a pseudoeffect expressed in formalisms which do not manifestly enforce causality \cite{Brodsky:1997de}, such as instant-form (IF) dynamics, a commonly used approach for the quantization of field theories.
In contrast, the explicitly causality-enforcing front-form (FF) framework of Dirac \cite{Dirac:1949cp}, has an essentially trivial vacuum, \textit{i.e}., a structureless Fock space ground state \cite{Casher:1974xd}.

A question thus arises: Does acceleration complexify the FF vacuum, so that the Unruh effect is also a feature of the FF?
If not, then the Unruh effect is itself a pseudoeffect, arising only in approaches that violate causality at intermediate stages of a calculation.
This conclusion would have profound implications, since
by Einstein's equivalence principle between gravity and acceleration \cite{einstein1907a}, the Unruh and Hawking effects~\cite{Hawking:1975vcx} are  typically considered to be related.
Alternatively, if the Unruh effect occurs in FF dynamics, then the usual interpretation of Unruh and Hawking phenomena, based on particles created from of a supposedly complex quantum vacuum,
is mistaken, because the effects would also emerge with a trivial vacuum.

Herein, we discuss the following three distinct cases:
(1) IF Minkowski space (\textit{i.e}., inertial) vacuum, which is complex;
(2) FF Minkowski space vacuum, which is trivial -- see Section~\ref{vacuum and commutation relations} and Appendix~\ref{FF-vacuum};
(3) ``Rindler vacuum'', associated with an accelerating frame for which ``Rindler coordinates'' are defined -- see Appendix~\ref{standard derivations}.
The Rindler vacuum is complex because Rindler fields are quantized at equal Rindler time.
Importantly, the three vacua are inequivalent because IF, FF and Rindler frames are not related by any Poincar\'e transformation.
Notably, although the Rindler vacuum is complex, it is, nevertheless, the vacuum for a Rindler observer; hence, imperceptible by definition.
In this connection, the Unruh effect is a statement that a Rindler (\textit{viz}. accelerated) observer perceives pairs that constitute the IF Minkowski space vacuum.

In section~\ref{Dirac's Forms of Dynamics}, we summarize important features of Dirac's three forms of dynamics \cite{Dirac:1949cp}
and, drawing on an analogy with classical Newtonian dynamics, explain the emergence and character of pseudoeffects in the IF formalism.
In Section~\ref{t-dependent boosts}, we explain why an Unruh effect is predicted when fields are quantized using the common IF approach, whereas that is not so when fields viewed from the inertial frame are quantized using the FF. 
The connection between a complex vacuum and causality violation in IF dynamics is elucidated in Section~\ref{vacuum and commutation relations} on commutation relations.
In Section~\ref{Gedanken experiment}, we discuss how the IF and FF findings can be reconciled.  This must be possible because natural (observable) phenomena do not depend upon the approach used to explain them.
Section~\ref{Conclusion} supplies concluding remarks.
%

\begin{figure*}[t]
 \center
\includegraphics[width=0.8\textwidth]{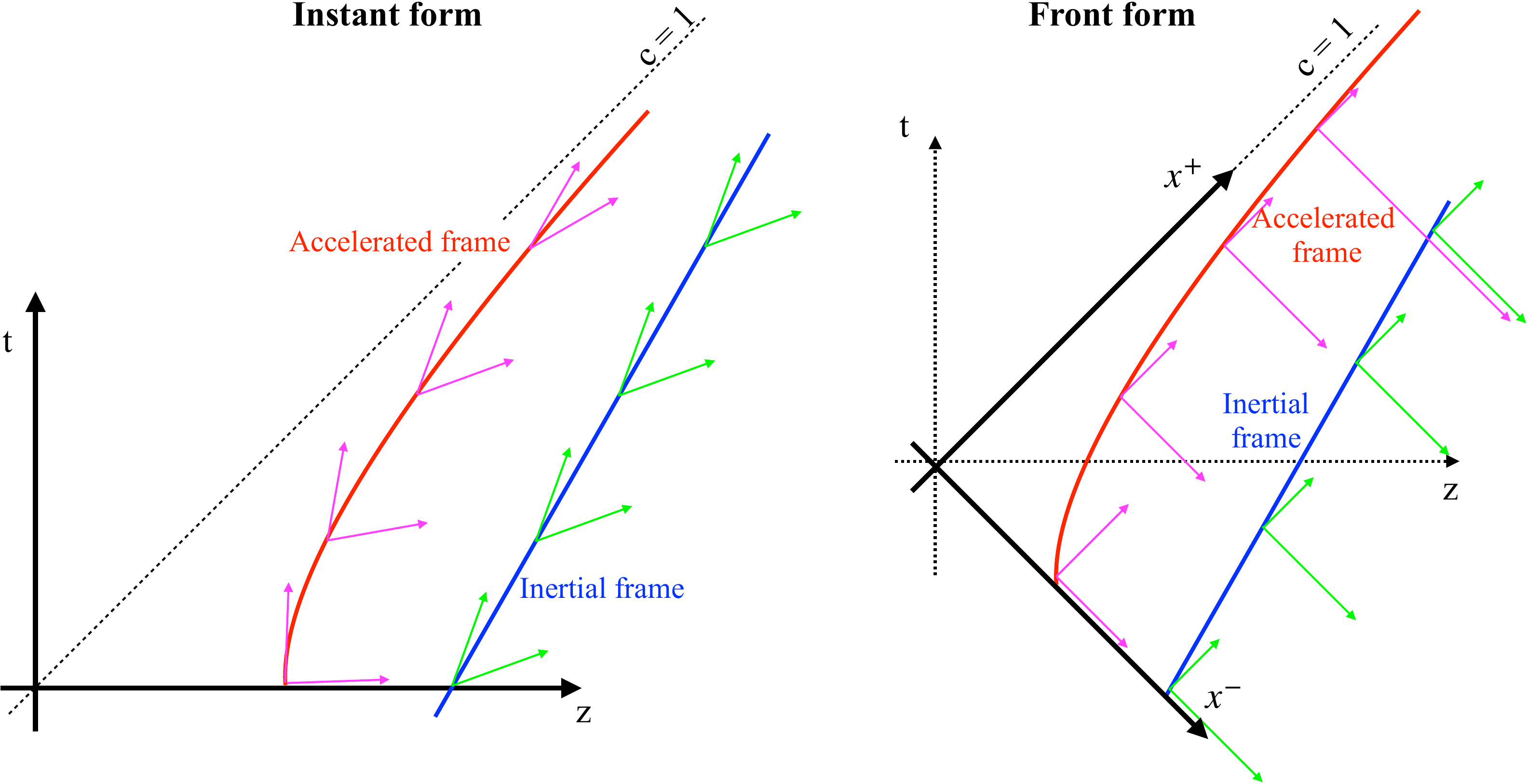}
\caption{Effect of boosts (blue worldline) and constant accelerations (red worldline) on attached IF (left) and FF (right) frames.
Thick arrows show the frame axes of an observer at rest.
The 45$^\circ$ dashed line is the trajectory for a particle moving at light speed ($c\equiv 1$ in natural units).
The $(-,+,+,+)$ metric is used in this figure, but $(+,-,-,-)$ is used for FF in the main text, following usage in particle physics where FF dynamics is employed. In that case, $x^-$ has the opposite direction, pointing left and upward.
\label{fig:IF_and_FF}
}
\end{figure*}

\section{Forms of Dynamics\label{Dirac's Forms of Dynamics}}
There is no unique choice of time parameter in relativistic kinematics. Parametrizing the worldline amounts to a specific convention for foliating the four-dimensional (4D) spacetime into 3D space + 1D time.
There are three pertinent conventions (forms) fulfilling relativistic invariance, basic causality, and fundamental spacetime symmetry requirements \cite{Dirac:1949cp}:
IF, where time is the usual Galilean time
(Fig.~\ref{fig:IF_and_FF}, left panel);
FF, where time is aligned tangentially to the light-cone (Fig.~\ref{fig:IF_and_FF}, right panel);
and Point-Form (PF), where the equal time hypersurfaces are hyperboloids.
For future reference, the FF coordinates are 
\begin{eqnarray}
\label{eq:FF-coordinates}
\tau \equiv x^+ := t+z ~~~{\rm(time)}\,; x^- := t-z ~~~{\rm(space)};\\
p^- := \omega - p ~~~{\rm(energy)};  p^+  := \omega + p~~~{\rm(momentum)}.
\end{eqnarray}
The labels here highlight the FF associations, and ($\omega ,p$) is the IF energy-momentum.
Importantly, distinct forms of dynamics are not related by Poincar\'e transformations.
Furthermore, they are not equivalent: each form of dynamics possesses a different stability group, \textit{viz}.\ a distinct set of Poincar\'e generators that are purely kinematical.
Such generators operate in constant form-time hypersurfaces, whereas dynamical operators evolve the system from one hypersurface to the next.

The Poincar\'e group has 10 generators:
4 linear momenta for spacetime translations;
3 angular momenta for space rotations;
and 3 Lorentz boosts.
The FF stability group is the largest, with 7 generators:
the linear momenta, the boosts and one angular momentum.
The IF and PF stability groups have only 6 generators:
IF -- linear and angular momenta;
and PF -- boosts and angular momenta.

Besides FF's larger stability group, crucial advantages are offered by the kinematical character of its boosts \cite{Brodsky:2022fqy}.
For instance, contrary to IF dynamics, popular in textbooks, and to other forms of dynamics, such as PF and oblique FF \cite{Das:2005gm}, FF dynamics manifestly preserves causality and the number of particles in systems related by boosts.
These things are key to its success in proving exact results in high-energy physics -- see, \textit{e.g}., Ref.\,\cite{Lepage:1980fj}.

In stark contrast, IF does not straightforwardly respect causality because it allows for vacuum-induced particle pairs, whose creation events are uncorrelated with the worldline of the system; hence, acausal.
Furthermore, the dynamics induced when IF-boosting a system often complicates its description with pseudoeffects \cite{Brodsky:2022fqy}.
As elaborated in Section~\ref{vacuum and commutation relations}, since commutation relations are imposed at equal form-time in canonical quantization
\textit{i.e.}, along the light-cone in the FF case, the FF vacuum is trivial.
(The possible existence of zero-modes \cite{Brodsky:1997de} is irrelevant to the Unruh effect -- see Appendix~\ref{FF-vacuum}.)

The conclusion reached in this article is that the Unruh effect is a pseudoeffect. 
Before demonstrating that, it is useful to elucidate the meaning and significance of this conclusion using an analogy from non-relativistic 
mechanics.

The symmetry of spacetime in the non-relativistic limit is Galilean invariance.
Analyses performed in a frame that violates this basic symmetry, {\it i.e.}, a non-Galilean (non-inertial) frame induce pseudoeffects, such as Coriolis, centrifugal and Euler forces. 
In the relativistic case, the spacetime symmetry is Poincar\'e invariance and analyses in frameworks that violate it likewise generate pseudoeffects. 
Once spacetime is foliated into space and time, preserving Poincar\'e invariance in the foliated space requires that there exists a single dynamical operator (the time-evolution operator, typically the Hamiltonian), with all other operators being kinematical, otherwise they would also evolve the system. 
Plainly, since the Hamiltonian is the quantity that contains the fundamental interactions, evolution generated by other operators is merely an expression of pseudodynamics. 

Dirac showed that pseudodynamics arise for the IF foliation because, in addition to the Hamiltonian, the three boost operators are dynamical \cite{Dirac:1949cp}. 
Thus, pseudoeffects appear each time the description of a system requires a change of frame, {\it e.g.}, in particle physics, where particles are frequently boosted from one frame to another and, even more clearly, with the Unruh effect, where several frames enter its description.

However, Dirac also showed that for the FF foliation, there are only 2 dynamical operators besides the Hamiltonian
, and that these correspond to two space rotations. 
The rotation axes are transverse to the spatial coordinate $z$ used to define the FF coordinates, \eqref{eq:FF-coordinates}, typically the direction of the particle motion. 
The two rotations are irrelevant in particle physics problems where the only rotational symmetry is around the displacement axis. 
Thus, FF respects Poincar\'e invariance for particle physics studies. Likewise, rotational invariance in the transverse directions is unconnected with the Unruh problem, and FF also respects Poincar\'e invariance in that case.

The analogy between the Unruh effect and classical pseudodynamics is now apparent. 
Nevertheless, a few points may benefit from elaboration.
Just as the derivation of, {\it e.g.}, the centrifugal force in a non-Galilean frame is valid, the derivation of the Unruh effect in IF is valid. 
We do not suggest otherwise. 
Rather, the central issue is that the effect originates from pseudodynamics and is therefore neither an objective nor fundamental part of quantum field theory (QFT).
The Unruh effect is necessary for the consistency of any IF analysis and it survives if that analysis is incomplete.
Drawing again on the non-relativistic analogy, just like overlooking a pseudoforce ({\it e.g.}, the Coriolis force) would lead to an incorrect conclusion, not accounting for the Unruh effect in an IF analysis will lead to internal inconsistencies. 
A corollary is that {\it all} pseudodynamics must be accounted for, which is a non-trivial requirement.
Just like analyses in Galilean frames avoid these complications, analyses in the Poincar\'e-preserving FF do, too.
A FF analysis, complete by Dirac's construction, shows that the Unruh effect is not observable because, in contrast to the aforementioned analogy, where non-relativistic pseudoeffects can be experienced by attaching oneself to a non-Galilean frame, the IF or FF frameworks have no material realization. 
The perception that the Unruh effect is observable thus originates from an incomplete IF analysis. 

In Section~\ref{Gedanken experiment}, we discuss another pseudoeffect that balances the Unruh effect, ultimately reconciling the IF and FF conclusions. 
Again, our arguments do not entail that IF analyses which produce an Unruh effect are wrong, only that such analyses are incomplete. They have omitted a subjective pseudodynamical effect whose contribution is required when one adopts a  choice of analysis framework that does not preserve relevant symmetries at intermediate stages of a calculation.

\section{Unruh effect\label{t-dependent boosts}}
Basic discussions of an Unruh effect focus on a free scalar field in (1+1) spacetime dimensions. Here, we also follow this simplification. 

The Unruh effect is predicted in IF dynamics because in non-inertial IF frames a field $\phi$ cannot unambiguously be factored into distinct time-dependent and space-dependent parts, namely, separated in a Poincar\'e-invariant manner \cite{Carroll:2004st}.
Consequently, in the decomposition of $\phi$ into
what may, for ease of discussion, be called positive and negative frequency modes, $f_p$, $f_p^*$:
\begin{equation}
\label{eq:field decomposition-1}
\phi = \int dp \,\big( \hat a_p f_p + \hat a_p^\dag f_p^*  \big),
\end{equation}
the assignment of positive or negative frequency mode is frame dependent.
($\hat a_p$, $\hat a_p^\dagger$ are, respectively, annihilation, creation operators.  $f_p$ is a mode. A positive (resp.\ negative) frequency mode is defined as  
$\partial_t f_p$=$-i \omega f_p$ (resp.\ $\partial_t f_p$=$~i \omega f_p$), 
with $t$ the proper time, a Minkowski metric sign (-.+,+,+) and the frequencies being always positive, $\omega \geq 0$.)

In another frame, $\phi = \int dp' \big( \hat b_{p'} f_{p'} + \hat b_{p'}^\dag f_{p'}^*  \big) $, with $\hat b_{p'} = \alpha \hat a_p + \beta \hat a_p^\dag $: $\beta \neq 0$ if one of the IF frames is non-inertial.
A Bogolyubov transformation \cite{Bogolyubov:1958km}
expresses this mixing:

\begin{eqnarray}
\label{eq:Bogolyubov transformation}
\begin{bmatrix} \hat b_{p'} \\  \hat b_{p'}^\dag \end{bmatrix}  =
\begin{bmatrix} \alpha_{11} & \beta_{12} \\ \beta_{21} & \alpha_{22} \end{bmatrix}
\begin{bmatrix} \hat a_p \\  \hat a_p^\dag \end{bmatrix} .
\end{eqnarray}
Since $\hat a_p^\dag$ and $\hat b_{p'}^\dag$ are creation operators, the vacuum of frame 1, satisfying $\hat a_{p} \ket{0}_1\equiv0$ by definition,
is perceived from frame 2 as containing particles: $\hat b_{p'} \ket{0}_1$$= (\alpha_{11} \hat a_p + \beta_{12} \hat a_p^\dag ) \ket{0}_1$$=\beta_{12} \ket{p} $$ \neq 0$.

Typical formal analyses leading to a prediction of the Unruh effect, involving Rindler frames, are provided in Appendices~\ref{standard derivations} and \ref{Unruh and thermal QFT}.
Here, we approach the problem by beginning with a boost and observing the effect of making its rapidity time-dependent, so that the boosted frame accelerates.
We perform the analysis
for IF and FF dynamics in parallel, using normal fonts for IF-specific formulae and {\bf bold fonts} for their {\bf FF equivalents}.
The IF analysis employs the {\it conventional relativistic} metric ${ (-,+,+,+)}$, with which the Unruh effect is usually discussed,
whereas the FF derivation follows the {\it particle physics convention}  $\bm{ (+,-,-,-)}$.
The latter implies the FF product
$\bm{a^\mu b_\mu}$ $\bm{=}$ $\bm{(1/2)(a^+b^- + a^- b^+)}$ in  (1+1)D spacetime.
%
%
Given the metrics, the positive frequency modes are  (recall $\tau = x^+$):
\begin{equation}
\label{eq: Minkowski IF and FF modes}
{ f_p \propto  e^{i (\omega t -  p z)} ~{\rm (IF) }}{\rm ;}
\quad
\bm{  f_{p^+} \propto  e^{-\sfrac{i}{2}(p^-\tau + p^+x^-)} ~{\rm (FF)}}.
\end{equation}

For pedagogy, we first discuss the case of inertial frames and then proceed to the accelerated frame case.

\subsection{Inertial frame case\label{Boost-case}}
It is straightforward to show that a positive (or negative) frequency mode of a wave expressed in an inertial frame remains so in another inertial  frame \textit{viz}.\ one related to the first frame by a Lorentz boost.
Figure\,\ref{fig:IF_and_FF} shows the effect of the boost (blue straight worldline with green coordinate axes):
in IF dynamics, it generates a rotation of the coordinate axes through an angle determined by $\theta$, the rapidity,  whereas the orientation of the axes is unaffected in FF. 

The Lorentz boost transformation formulae in the IF and FF frameworks are:
{\allowdisplaybreaks
\begin{equation}
\label{eq:Lorentz transfo for spacetime}
\begin{array}{llll}
  t^\prime     =    t \cosh\theta    -    z \sinh\theta & ({\rm IF}) ;
& ~ \pmb{ \tau^\prime    =    {\rm e}^\theta\tau} & ({\rm FF}),\\
  z^\prime    =    z \cosh\theta    -    t \sinh\theta &  ({\rm IF}) ; 
& ~ \pmb{ x^{-\prime}    =    {\rm e}^{-\theta} x^- } & ({\rm FF}),
\end{array} 
\end{equation}
or, for energy-momentum:
\renewcommand{\arraystretch}{1.}
\begin{equation}
 \label{eq:Lorentz transfo for energy-momentum}
 \begin{array}{llll}
      \omega^\prime    =    \omega \cosh \theta    -    p \sinh \theta &  ({\rm IF});
      &  \pmb{ p^{+\prime}    =    {\rm e}^{\theta} p^+} & ({\rm FF}),\\
      p^\prime    =    p \cosh \theta    -    \omega \sinh \theta &  ({\rm IF});
      &  \pmb{ p^{-\prime}    =    {\rm e}^{-\theta} p^-} & ({\rm FF}).
 \end{array}
\end{equation}
The inverse transforms are straightforwardly obtained by changing $\theta \to -\theta$ and switching $t^\prime\leftrightarrow t$}, etc.:
\begin{equation}
 \label{eq:Lorentz inv. transfo}
 \begin{array}{llll}
t     =    t^\prime \cosh\theta + z^\prime \sinh\theta  & ({\rm IF}) ;
&\pmb{ \tau    =    {\rm e}^{-\theta}\tau^\prime} & ({\rm FF}),\\
 z    =    z^\prime \cosh\theta + t^\prime \sinh\theta  &  ({\rm IF}) ;
& \pmb{ x^{-}    =    {\rm e}^{\theta} x^{-\prime} } & ({\rm FF}).
 \end{array}
\end{equation}
In the boosted frame, the sign of a mode frequency can be identified via its time derivative: 
\begin{equation}
 \label{eq:time-dep mode Lorentz transfo-1}
\begin{array}{llll}
\displaystyle
\frac{\partial f_p}{\partial t^\prime }
 \hspace{-0.8mm} = \hspace{-0.8mm}  \frac{\partial x^\mu}{\partial t^\prime} \partial_\mu f_p
 \hspace{-0.8mm} = \hspace{-1.5mm}  \left[
\frac{\partial t}{\partial t^\prime} \partial_t  \hspace{-0.5mm} + \hspace{-0.5mm}  \frac{\partial z}{\partial t^\prime} \partial_z
 \hspace{-0.5mm}  \right]  \hspace{-1mm}  f_p   ({\rm IF});
     \displaystyle
    \pmb{ \frac{\partial f_{p^+}}{\partial \tau^\prime }  \hspace{-0.8mm} = \hspace{-0.8mm} 
    \frac{\partial x^\mu}{\partial \tau^\prime } \partial_\mu f_{p^+}  \hspace{-0.8mm} = \hspace{-1.5mm} 
    \left[
    \frac{\partial \tau}{\partial \tau^\prime} \partial_\tau  \hspace{-0.8mm} + \hspace{-0.8mm}  \frac{\partial x^-}{\partial \tau^\prime } \partial_{x^-}
   \hspace{-0.8mm}   \right]  \hspace{-1mm}  f_{p^+}}  ({\rm FF}).
\end{array}
\end{equation}
{\allowdisplaybreaks
Using Eqs.~(\ref{eq: Minkowski IF and FF modes}) and~(\ref{eq:Lorentz inv. transfo}), this becomes:
\begin{equation}
 \label{eq:time-dep mode Lorentz transfo-2bis}
\begin{array}{lllll}
\displaystyle
 \frac{\partial f_p}{\partial t^\prime }
=
i \left[\omega \cosh\theta -  p \sinh\theta \right] f_p  & ({\rm IF}) \,;~~~
      \displaystyle
    \pmb{ \frac{\partial f_{p^+}}{\partial \tau^\prime } =
     -e^{-\theta} \frac{i}{2}p^- f_{p^+}} & ({\rm FF})\,,
\end{array}
\end{equation}
or, using Eq.~(\ref{eq:Lorentz transfo for energy-momentum}),
\begin{equation}
 \label{eq:time-dep mode Lorentz transfo-2}
\begin{array}{lllll}
\displaystyle
\frac{\partial f_p(x^\mu,p^\mu)}{\partial t^\prime }
= i \omega^\prime f_p (x^\mu,p^\mu) & ({\rm IF}) \,; ~~~
 \displaystyle
    \pmb{\frac{\partial f_{p^+}(x^\mu,p^\mu)}{\partial \tau^\prime } =
    -\frac{i}{2}p^{-\prime}f_{p^+}(x^\mu,p^\mu)} & ({\rm FF})\,,
\end{array}
\end{equation}
\textit{i.e}., the frequency in a boosted frame is the boosted frequency, $\omega^\prime$ or $\pmb{p^{-\prime}}$.
The same holds for negative frequency modes, $f_p^* $ and $f_{p^+}^*$.
Using IF, the dispersion relation $\omega = |k|$ implies $\omega' = \omega (\cosh \theta \pm \sinh \theta) = \omega e^{\pm \theta}$. With FF, one has $\pmb{p^{-\prime} = p^{-}e^{-\theta} }$
(Eq.~\eqref{eq:Lorentz transfo for energy-momentum}).
An exponential function is always positive, so the signs of the mode derivatives are consistent in the initial and boosted frames.
Thus, for a time-independent boost, there is no mixing of positive and negative frequency modes;
the Bogolyubov transform is diagonal;
and the vacua of the original and boosted inertial frames coincide.

\subsection{Accelerated frame case\label{Accel-case}}
We now consider transformation to an accelerated frame.
The latter can be formalized by making $\theta$ vary with time -- see Fig.\,\ref{fig:IF_and_FF}.  We set $\theta=ut$ or $\pmb{ \theta=\upsilon\tau}$, with $u\neq 0 \neq \pmb{\upsilon}$ and both small, so that acceleration is constant to a good approximation.
Importantly, $\theta$ is not explicitly space-dependent:
\begin{equation}
 \frac{\partial \theta}{\partial x}=0    \quad ({\rm IF}),
\qquad \pmb{  \frac{\partial \theta}{\partial x^-}=0} \quad ({\rm FF}).
\end{equation}
Using IF dynamics and the appropriate entries in Eqs.~(\ref{eq:Lorentz transfo for spacetime}--\ref{eq:Lorentz inv. transfo}), 
then \eqref{eq:time-dep mode Lorentz transfo-1} yields:
\begin{align}
\nonumber
\frac{\partial f_p}{\partial t^\prime } & = 
\big[
\big(
[1+z^\prime \partial_{t^\prime}\theta]\cosh\theta
+t^\prime \partial_{t^\prime}\theta \sinh\theta \big) \partial_t
+  
 \big(
[1+z^\prime\partial_{t^\prime}\theta] \sinh\theta 
+ t^\prime\partial_{t^\prime}\theta \cosh\theta 
\big)\partial_z
\big] f_p \nonumber \\
  &=  i \left[ \omega^\prime + \partial_{t^\prime} \theta (\omega z - t p) 
\right]f_p\, ; 
 \label{eq:time-dep mode accel. transfo-1 IF}
\end{align}
whereas, using the FF,
\begin{align}
\pmb{  \frac{\partial f_{p^+}}{\partial \tau^\prime } }
\pmb{ 
 \hspace{-0.8mm} = \hspace{-0.8mm}  \left[
{\rm e}^{-\theta}
( 1 - \tau^\prime \partial_{\tau^\prime}\theta) \partial_\tau
+ {\rm e}^\theta \partial_{\tau^\prime} \theta x^{-\prime} \partial_{x^-}
\right] f_{p^+}}  = \pmb{ 
-\frac{i}{2}\left[
p^{-\prime} + \partial_{\tau^\prime}\theta (x^- p^+ - \tau p^- ) \right] f_{p^+}\,.
}
\label{eq:time-dep mode accel. transfo-2}
\end{align}
Evidently, the IF and FF expressions are similar:
\eqref{eq:time-dep mode accel. transfo-1 IF} \textit{cf}.\
\eqref{eq:time-dep mode accel. transfo-2}.
Moreover, reviewing Eqs.~(\ref{eq:Lorentz transfo for spacetime}--\ref{eq:Lorentz inv. transfo}),
one sees that
$\omega z-tp$ = $\omega^\prime z^\prime - t^\prime p^\prime$ and $\pmb{ x^-p^+ - \tau p^-}$ = $\pmb{ {x^{- \prime}p^{+\prime} - \tau^\prime p^{-\prime}}}$; 
the combinations are Poincar\'e invariant scalar products, which we write as
$\tilde p^\mu x_\mu = p^\mu \tilde x_\mu$:
\begin{equation}
\begin{array}{llll}
\displaystyle  \frac{\partial f_p}{\partial t^\prime } =
i \left[ \omega^\prime + p^\mu \tilde x_\mu \partial_{t^\prime}\theta \right] f_p 
      & ({\rm IF})\,; ~~~
 \displaystyle \pmb{ \frac{\partial f_{p^+}}{\partial \tau^\prime } =
-\tfrac{i}{2} \left[  p^{-\prime} + p^\mu \tilde x_\mu \partial_{\tau^\prime}\theta \right] f_{p^+}}
      & ({\rm FF})\,.
\end{array}
\label{eq:time-dep mode accel. transfo-3bis}
\end{equation}

Compared with the inertial case, \eqref{eq:time-dep mode Lorentz transfo-2bis}, a Poincar\'e-violating term  $\partial_{t^\prime}$ or \pmb{  $ \partial_{\tau^\prime}$} has appeared. Of course, the appearance of that derivative means the new term violates Poincar\'e invariance, as was to be expected since inertial and accelerated frames are not related by genuine boosts, for which $\theta = \,$constant.

Considering the IF case, the time direction changes along the accelerated worldline; so, time and space mix -- see Fig.~\ref{fig:IF_and_FF}.
This is expressed in the result
\begin{eqnarray}
\partial_{t^\prime}\theta(t)
= u \cosh(u t^\prime) [1 + u z^\prime + u t^\prime \tanh(u t^\prime)] 
= u \, {\rm sech}(u t)/[1 - u z + u t \tanh (u t)].
\label{eq:rapidity-derivative-IF}
\end{eqnarray}
The variation of the sign of $\partial f_p / \partial t^\prime$ makes evident that one cannot 
develop a Poincar\'e-invariant separation of the field $\phi$ into time and space components. 
Since $p^\mu \tilde x_\mu=\omega(z\pm t)\geq 0$, the sign of $\partial f_p/\partial t^\prime$ is determined by $\partial_{t'} \theta$,  \eqref{eq:rapidity-derivative-IF}. Fig.\,\ref{fig:rapidity-derivative} reveals its rapid oscillation with $t$ and $z$ (or $t'$ and $z'$). 
So, what were positive and negative frequency modes in the IF inertial frame mix in the IF accelerated frame: the Bogolyubov matrix is not diagonal and there is an Unruh effect in this development of the IF case.

\begin{figure*}[t]
 \center
\includegraphics[width=0.49\textwidth]{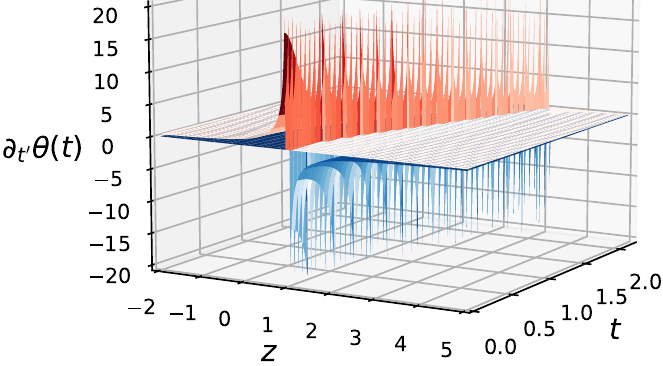}
\hspace{-5mm}
\includegraphics[width=0.49\textwidth]{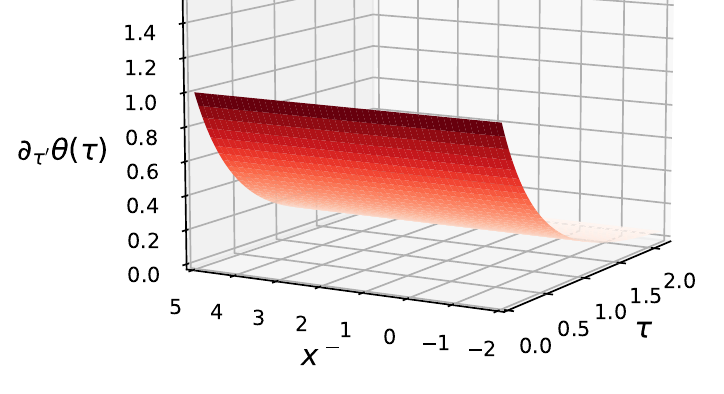}
\caption{
Derivative of the accelerated frame rapidity with proper time.
Left panel: IF case $\partial_{t^\prime}\theta(t)$.  
Right panel: FF case $\partial_{\tau^\prime}\theta(\tau)$.  
Red indicates positive values and blue, negative values.  
The swiftly oscillating sign in the IF case reveals the Unruh effect: the inertial frame's positive and negative frequency modes become mixed in the IF accelerated frame. The discrepancy in mode definitions between the IF inertial frame and the accelerated frame entails a non-diagonal Bogolyubov matrix; hence, an Unruh effect. 
In the FF case, time and space do not mix: $\partial_{\tau^\prime}\theta(\tau)$ is independent of $x^-$. 
Consequently, the sign of $\partial_{\tau^\prime}\theta(\tau)$ stays the same, allowing for a consistent definition of positive and negative frequency modes between the FF inertial and accelerated frames. The Bogolyubov matrix is diagonal, precluding an Unruh effect in the FF framework.
\label{fig:rapidity-derivative}
}
\end{figure*}

Using FF, on the other hand, the time direction remains fixed along the accelerated worldline -- see Fig.~\ref{fig:IF_and_FF}.
Thus, the time-derivative of $\theta$ is solely a function of time: 
\begin{eqnarray}
\pmb{
\partial_{\tau^\prime}\theta(\tau) = \upsilon {\rm e}^{- \upsilon \tau^\prime} [1 - \upsilon \tau^\prime]
= \upsilon {\rm e}^{- \upsilon \tau}/[1 + \upsilon \tau]
}.
\label{eq:rapidity-derivative-FF}
\end{eqnarray}
Consequently, with time and space directions remaining mutually perpendicular, $\phi$ can still be unambiguously expanded over positive and negative frequency modes, even for accelerated frames.
Indeed, the sign of $\bm{\partial f_{p^+} / \partial \tau^\prime}$ is determined by $\bm{\partial_{t'} \theta}$,  \eqref{eq:rapidity-derivative-FF}, which sign
is constant -- see Fig.\,\ref{fig:rapidity-derivative}, right panel. Therefore, the definition of positive and negative frequency modes in the FF inertial frame and accelerated frame are consistent,
the Bogolyubov matrix remains diagonal, and the vacua of frames coincide and are trivial:
 {\it there is no Unruh effect in FF dynamics}.

There is a straightforward explanation for arriving at a prediction of the Unruh effect when using IF dynamics but not with the FF.
Namely, boost operators are dynamical in the IF -- see Section~\ref{Dirac's Forms of Dynamics}: they mix kinematics and dynamics, leading to non-conservation of particle number.
In an accelerating frame, this mixing changes with time.
Thus, non-stationary dynamics emerges with energy being transferred between systems, which would be an observable effect, unless cancelled by some other process.
In contrast, FF boosts are kinematical operators.
Hence, no dynamical effects are introduced by acceleration. So, an Unruh detector remains quiet.

\section{Connection to commutation relations\label{vacuum and commutation relations}}
Canonical quantization of fields is performed at constant proper time, \textit{i.e}.,
with fields defined along the fixed $z$-direction for the IF inertial frame
(Fig.~\ref{fig:IF_and_FF}, left panel),
along the varying $z$-direction for the IF accelerated frame,
and along the fixed $45^\circ$ $x^-$-direction for both inertial and accelerated FF frames (Fig.~\ref{fig:IF_and_FF}, right panel).

 \begin{figure*}[t]
 \center
\includegraphics[width=1\textwidth]{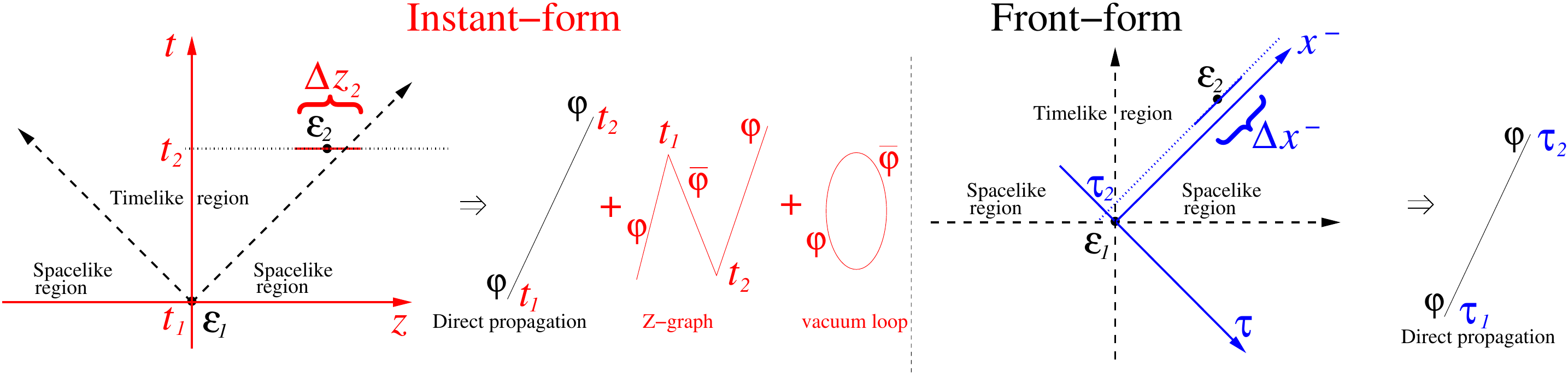}
\caption{Origin of IF vacuum complexity.
A field $\varphi$ propagates between events $\mathcal{E}_1$ and $\mathcal{E}_2$.
Equal-time commutation relations imply uncertainty relations along the fixed-time hypersurface (dotted line).
In IF dynamics, timelike events with respect to $\mathcal{E}_1$ may become spacelike owing to the Heisenberg uncertainty $\Delta z_{_2}$
(left panel).
Thus, the time-ordering of $\mathcal{E}_1$ and $\mathcal{E}_2$ is frame-dependent.
When $t_2<t_1$, causality-violating $Z$-graphs arise.
Their negative probabilities are compensated by vacuum loops, rendering the vacuum complex.
The loops' virtual particles may borrow 4-momentum from an accelerated frame and become observable in that frame (Unruh effect).
In FF dynamics, commutation relations are set at fixed FF time $\tau$.
Thus, the uncertainty $\Delta x^-$ never accesses the spacelike region: $\mathcal{E}_2$ is always timelike with respect to $\mathcal{E}_1$ (right panel). No $Z$-graphs arise, virtual loops are forbidden by unitarity, and the vacuum is trivial.
Without vacuum loops, no Unruh effect emerges.}
\label{fig:Heisenberg}
\end{figure*}

In canonical quantization, the Heisenberg uncertainty principle originates from commutation relations \cite{Kennard:1927vgq}.
Since they are imposed at equal time, the uncertainty principle also operates
on equal time hypersurfaces.
This results in a complex vacuum when time is defined as IF time
because events that should be causally linked, \textit{i.e.}, timelike-separated, may become spacelike-separated owing to position uncertainty.

Consider Fig.\,\ref{fig:Heisenberg}, which shows the propagation of a massive field
  $\phi$ between events $\mathcal E_1$ and $\mathcal E_2$ at times $t_1$ and $t_2$, whereat the field interacts with an external scalar potential $U$. In a classical treatment, the propagation is restricted to the timelike region, {\it viz.} the upper part of the light-cone. However, quantum fluctuations, intuitively visualized in Fig.\,\ref{fig:Heisenberg}-left as a Heisenberg uncertainty on the particle position at $t_2$, allow for instances of $\mathcal E_2$ to leak out of the light-cone. In that case,
$\mathcal E_1$ and $\mathcal E_2$ are spacelike-separated,  and the time-ordering of $\mathcal E_1$ and $\mathcal E_2$ is frame-dependent with $t_2 < t_1$ in some frames where they form $Z$-graphs with event $\mathcal E_2$ the spontaneous appearance of a particle-antiparticle pair.
The antiparticle annihilates at later time $t_1$ with the initial particle.
This description is acausal since the pair appearance is spontaneous and cannot be caused by the original particle propagation owing to the spacelike interval.

 More formally, the propagation amplitude for $\phi$ is $${\mathcal I}=\int dx_1 dx_2\int {\frac{dP}{(2\pi)^3 4E^2}}U^*(x_2)\phi^*( x_2)e^{-i P_\mu (x_2^\mu - x_1^\mu) }U( x_1)\phi( x_1),$$ with $P^\mu$ the field four-momentum and the energy sign fixed, {\it e.g.}, $P^0\geq0$. This last point is crucial because a function that is Fourier decomposed with only positive (or only negative) energies cannot be zero for a finite $t$ range, unless it vanishes everywhere. Therefore, ${\mathcal I}$ can be nonzero for spacelike $ x_2^\mu -  x_1^\mu$ and spacelike processes are allowed in IF. In frames with $t_2>t_1$, they appears as $Z$-graphs, which
contribute negative probabilities to the propagation, so disconnected creation-annihilation diagrams (vacuum loops), balancing the negative probabilities, must be introduced to preserve unitarity \cite{Feynman:1987gs}.
 The virtual particles in the loops can cause state $\ket i \to \ket j$ transitions in a detector, but such a transition is followed, after a time $\delta t \sim |E_i - E_j |^{-1}$, by an opposite $\ket j \to \ket i$ transition, resulting in no net excitation of the detector, as demanded in general by Lorentz invariance and, specifically, by the field correlation encoded in the propagator of that field.
In accelerated frames,   however, after $\delta t$, the energy gap $ |E_i - E_j |$ would have been boosted to a different value and the correlated $\ket j \to \ket i$ back transition no longer matches the boosted energy gap~\cite{Zee:2013dea}.
Thus, the virtual particles in the vacuum loops borrow 4-momentum from the acceleration process and, unless this is somehow cancelled, become  observable, leading to the Unruh effect.
An acceleration of a FF frame continuously rescales the axes (Fig.\,\ref{fig:IF_and_FF}, right panel) without reorienting them; hence, the uncertainty principle continues to operate along the light-cone direction $x^-$.
The events $\mathcal E_1$ and $\mathcal E_2$ remain timelike-separated, causality is preserved, and unitarity forbids vacuum loops \cite{Brodsky:2022fqy}.
Vacuum simplicity persists in an accelerated FF frame and, without vacuum loops in any type of FF frame, there is no Unruh effect.
Consistently, momentum conservation also prohibits vacuum loops:
FF particles must have non-negative momenta $p^+$ \cite{Brodsky:1997de, Casher:1974vf, Kogut:1974sn};
and since $p^+ = 0$ for the vacuum, by definition,~\cite{note-FFmom} 
%
one of the particles of the hypothetical vacuum loop would have $p^+ < 0$, which is forbidden.

\section{Reconciling Instant and Front Form Descriptions\label{Gedanken experiment}}
Physics is independent of any choice of frame or framework.
Therefore, IF and FF conclusions must finally agree.
Here, we describe how one may reconcile the apparently conflicting pictures drawn above.   A detailed quantitative discussion involves extensive calculation of nuclear structure. We provide, as an illustration, the outline of such a study.

Following Ref.\,\cite{Brodsky:1968xc}, which reconciled the IF and FF descriptions of deuteron structure, we consider an Unruh detector \cite{Unruh:1976db, Hawking:1979ig} made of an ensemble of particles consisting of two bound fermions.
Such a detector would monitor the distribution of the particle energy states.
The temperature $T$ is obtained from the population distributions:
$N_i=N \exp{(- E_i/[k_B T])}/Z$,
where $N$ is the total number of particles;
$E_i$, the energy of state $\ket{i}$,
$k_B$, Boltzmann's constant,
and $Z$, the partition function.
Irrespective of the Unruh effect, the detector {\it structure} in the IF case continuously changes with
time because of a combination of three facts:
(1) an IF boost is dynamical;
(2) the rapidity becomes time-dependent for accelerations, $\theta \to \theta(t)$;
and
(3) a boost is equivalent to a Rindler time translation -- see Appendix~\ref{Unruh and thermal QFT}.

Specifically,  in IF dynamics, the boost generator decomposes into a kinematical part and an interaction-dependent dynamical part, $\hat{K}_{\rm dyn} = \frac{1}{2}(\hat{ {r}}\hat{V} + \hat{V}\hat{ {r}})$, where $r$ is the distance between the two particles composing the bound system and $V$ is the binding potential. (We include a $\hat .$ to distinguish operator from classical quantity.) Since $\hat{V}$ does not commute with the internal Hamiltonian, $\hat{H}_{\rm int}$, describing the bound system in its center-of-mass, $[\hat{K}_{\rm dyn}, \hat{H}_{\rm int}] \neq 0$, an IF boost does not preserve internal energy eigenstates. If the bound system undergoes, as in Section~\ref{Accel-case}, a succession of boosts of rapidity $\theta = u t$, this non-covariance produces an effective Hamiltonian $\hat{H}_{\rm boost} \simeq u\hat{ {r}}\braket V$, where $\braket V$ is an overage over the binding potential weighted by the form factor of the system.

Again, a succession of boosts of changing rapidity is equivalent to an acceleration by $u$, the system
proper acceleration. The generated $\hat{H}_{\rm boost}$ drives coherent  $\Delta L = \pm 1$ internal transitions, {\it viz}.
reversible Rabi-like oscillations between states. In a non-ideal gas, interactions between the bound
systems cause these coherences to damp out at a rate $r_{\rm dc}$, which produces an irreversible transition
rate $R_{\rm boost} \simeq \frac{2|\kappa|^2}{r_{\rm dc}}$ between the bound system states. Here, $\kappa = u \bra E \hat{K}_{\rm dyn} \ket g $, with $\ket g$ and $\ket E$ 
being the ground and excited states of the bound system, respectively. (For the deuteron, they are the $s$-wave ground state and the $p$-wave continuum).

The total internal energy flux, $\Phi_{\rm tot}$, in the gas ensemble can be decomposed into three coupled components: an Unruh flux, $\Phi_{\rm Unr}$, an ``IF-boost'' flux, $\Phi_{\rm boost}$ and a cross flux, $\Phi_{\rm cross}$. The  $\Phi_{\rm Unr}$ and $\Phi_{\rm boost}$ fluxes only depend explicitly on the parameters characterizing their effects, {\it viz}. the acceleration, $u$, and the Unruh detector characteristics for the former and, for the latter, $u$, the deuteron structure, and the nuclear force potential. On the other hand, $\Phi_{\rm cross}$ combines both sets of parameters. The cross term produces a negative interference between the Unruh and IF-boost effect because the Unruh bath acts as an additional decoherence channel,  $r_{\rm dc} \to r_{\rm dc} + w^{\rm Unr}$, where $w^{\rm Unr}$ is the Unruh transition rate density, thereby suppressing the efficiency of boost heating, {\it i.e.}, effectively producing a cooling mechanism.
Unruh detectors are typically described in terms of energy levels without considering the dynamics of the internal detector structure responsible for
those levels.\footnote{This is highlighted by the fact that the masses of the particles constituting the Unruh detector are not mentioned in descriptions. Were the internal dynamical structure underlying the energy levels discussed, masses would appear, as they do when dynamics is described.} 
Without such consideration, one cannot account for the IF boost dynamical effect.
Overlooking  the IF-boost generated transitions and the associated cooling that emerges from them, there is no compensation mechanism for the Unruh effect; hence, it seems objectively observable.

In FF dynamics, none of this occurs because boosts are kinematical (no   IF-boost transitions and cooling) and the FF vacuum is trivial (no Unruh temperature).
Since physics is independent of choice of frame and form, then the two IF effects must compensate each other so that IF and FF agree.
%
It is now evident, however, that since boost-induced dynamics are overlooked in the usual IF analyses that lead to an Unruh effect, it has hitherto been erroneously identified as observable.

\section{Discussion and conclusion\label{Conclusion}}
The Unruh effect is a widely held to be a feature of QFT. 
It owes its importance mainly to its relation with Hawking radiation, via Einstein's equivalence principle between gravity and acceleration.
It is popularly claimed that the Unruh effect needs no empirical verification, beyond that involved in validating Poincar\'e-invariant QFT as a description of Nature~\cite{Crispino:2007eb, Fulling:2014wzx}.
We have exposed the flaw in that position: FF quantization is a powerful predictive tool in QFT; yet, this approach reveals that prediction of an Unruh effect is an erroneous consequence of causality-violating aspects of conventional IF dynamics.

The crucial point is that any apparent dynamics generated by kinematic transformations, here the Lorentz boosts, is pseudodynamics by definition. 
In IF, boosts have long been known to generate such effects, while they are purely kinematical in FF. 
Then, a natural question is whether there should also be a rotational Unruh effect generated by centripetal acceleration. 
Evidently, the naive answer from the standard approach is {\it yes}. 
However, IF rotations are purely kinematical operations and, indeed, even if the IF is used to study a purely rotational Unruh effect, one finds that the effect is absent \cite{Denardo:1978dj, Letaw:1979wy, note_rot_Unruh}. 
This accords with our present finding: in FF, boosts are kinematical, leading to no Unruh effect, and in IF, rotations are kinematical, leading to no rotational Unruh effect -- counterintuitively from the usual IF standpoint. (There is also no rotational Unruh effect with the FF because the two relevant rotations are also kinematical.) 
This strongly supports the conclusions arrived at herein.

Properly, the Unruh effect is cancelled in IF dynamics by another pseudoeffect, namely, cooling of the Unruh detector owing to dynamical evolution
under a succession of IF boosts.
Consequently, notwithstanding popular perspectives, one is left with a final point that should not be lightly dismissed; namely, neither the Unruh effect nor any of its supposed corollaries can be considered real without empirical verification. 

\paragraph{Acknowledgments}
We are grateful to G.~de T\'eramond for useful discussions. 
This material is based upon work supported by 
U.S.\ Department of Energy, Office of Science, Office of Nuclear Physics, contract DE-AC05-06OR23177 (AD);
U.S.\ Department of Energy contract DE-AC02-76SF00515 (SJB); 
National Natural Science Foundation of China, grant no.\ 12135007 (CDR);
and the U.~S.~National Science Foundation award no.\ 1847771 (BT).}

\pagebreak

\appendix

\noindent {\bf \Large Appendices}

\section{Front Form Vacuum\label{FF-vacuum}}
\noindent{\textit{No heating}}\,---\,%
That virtual particle loops are absent from the front form (FF) vacuum is firmly established.
Nevertheless, the role of zero-momentum modes in the FF vacuum is a topic of intense discussion, \textit{e.g}., such issues as whether, in flat spacetime, the vacuum energy can be renormalized away are widely canvassed \cite{Collins:2018aqt, Mannheim:2019lss}.
Though important in themselves, these questions are irrelevant to the Unruh effect because $p^+=0$ modes cannot transfer momentum (equally, kinetic energy) to the particles forming the Unruh detector/thermometer: zero-modes cannot heat thermometers.
\smallskip

\noindent{\textit{Event horizons do not split zero modes}}\,---\,
The popular interpretation of the Unruh and Hawking effects suggests an intuitive explanation of why zero-modes are irrelevant in this connection. 
In that interpretation, one of the particles of the vacuum loop falls beyond the event horizon (for the Unruh effect, the Killing horizons are the $x^+$ and $x^-$ axes); so, cannot annihilate with its partner. 
Such a process requires space separations between the two particles and evolution, while zero-modes have no space or time extension, making them irrelevant. 
Thus, either zero-modes are immaterial; 
or the vacuum pair interpretation is erroneous;
or it is instant form (IF) specific, hence can play no role in an objective explanation.

\smallskip

\noindent{\textit{Constraints from General Relativity}}\,---\, Considering that, by Einstein's equivalence principle, accelerated frames are equivalent to curved spacetime, one arrives at another argument for the irrelevance of zero-modes, as well as a new perspective on the Unruh effect.

Indeed, it is known that the definition of positive and negative frequency modes is ambiguous in a curved spacetime/non-inertial frame, in contrast to Minkowski spacetime.
This is the formal reason for the Unruh effect, see Section~\ref{t-dependent boosts}.
These features apparently contradict the basic principle of general relativity that one must tend to a Minkowski spacetime in the limit of vanishing distances.
In other words, when distance scales are negligible in comparison with the curvature radius, space appears flat, with a unique vacuum.

Naturally, naive and classical vacua have no special scales and one may, unhindered, reach the small distance limit.
However, in the IF framework, the vacuum is complex, with a characteristic distance scale set by the size of the vacuum loops.
Such a scale forbids one from achieving the small-distance limit, keeping the definition of positive and negative modes ambiguous, and allowing for an Unruh effect. This perspective makes the crucial role of a non-trivial vacuum clear~\cite{EP-TD}.
Conversely, the FF vacuum is trivial; hence, without distance scales that would prevent approach to the Minkowski limit.
So, positive and negative modes are well defined, precluding an Unruh effect.

This discussion also elucidates why zero-point energy cannot contribute to an Unruh effect; namely, by definition, it is associated with an operator expectation value at a single point \cite{Mannheim:2019lss}, hence, devoid of a characteristic distance scale.  

\noindent{\textit{Nonperturbative vacuum}}\,---\,
We note that studies of the vacuum in QFT often employ a perturbative formalism; thus, one can consider whether the nonperturbative FF vacuum could be nontrivial.
This is also irrelevant to our subject because no field coupling or other expansion parameter is active in predictions of the Unruh effect, which is derived from kinematical considerations alone.
If such were not the case, then one could always choose a small coupling for which a perturbative treatment is sufficient, since typical analyses find the Unruh effect to occur universally, irrespective of coupling strength.
Additionally, discussions of the Unruh effect are often set for simplicity in (1+1)D \cite{Crispino:2007eb, Carroll:2004st}, where the triviality of the nonperturbative FF vacuum is clear: quantum chromodynamics has been solved exactly in (1+1)D, with the meson spectrum recovered in the number of colors $N_c \to \infty$ limit using FF \cite{tHooft:1973alw} or IF \cite{Bars:1977ud}.
Both studies involve obtaining the Hartree-Fock solution for its vacuum, but with the FF calculation being considerably simpler owing to vacuum simplicity.
Finally, vacuum structure, whether trivial or complex, is not invoked: the demonstrations only use coordinate definitions of the forms of dynamics, the consequent quantization conditions, and generic properties of QFT.
Only the interpretation of the result invokes the vacuum (perturbative) structure as it provides a seemingly intuitive picture.

\smallskip

\noindent{\textit{Infinite momentum frame}}\,---\,
Often, FF results are formally recovered by boosting IF results to the infinite momentum frame (IMF) \cite{Weinberg:1966jm}.
This is also the case here, since in the IMF, the Rindler, IF and FF times coincide -- see Fig.~1.
From Ref.\,\cite{Chmielowski:1994gb}, this implies the same vacua and, in turn, no Unruh effect.
Yet, one also needs to allow for the possibility of zero-modes in the IF-IMF vacuum.
Some of these modes correspond to real particles -- rather than the null-momentum vacuum loops (``tadpole graphs'') responsible for zero-point energy -- and might contribute to the static properties of bound states \cite{Ji:2020baz}.
However, consistent with the FF result, even these particular IF-IMF zero-modes cannot lead to the dynamics underlying the Unruh effect for several reasons.
First, such modes manifest themselves only within bound states, potentially contributing inside to long-distance field correlations (sometimes called ``condensates'') without support outside a bound state \cite{Brodsky:2009zd, Brodsky:2022fqy, Brodsky:2010xf, Brodsky:2012ku}.
Second, IF-IMF vacuum loops or condensates cannot produce pairs of physical particles for the same reason as for the FF 
(see Section~\ref{vacuum and commutation relations}): 
all particles must have nonnegative IF-IMF momenta.
Since the vacuum or zero-modes have null momenta by definition, and since momentum is conserved, this excludes loops because one of the particles would have a negative momentum.
Considering the IMF limit offers a simple interpretation of the absence of the Unruh effect: in the IMF the acceleration must fall to zero; thus, so must the Unruh temperature.

\smallskip

\noindent{\textit{Summary}}\,---\,
Herein, FF vacuum triviality refers to the absence of loops of particles with non-zero momenta, without considering zero-modes, which might play a role in some physics but are irrelevant to the Unruh effect.
Furthermore, vacuum triviality is not employed in the FF analyses and Appendices~\ref{standard derivations}--\ref{Dynamical derivation}:
an Unruh effect is precluded in FF quantization for the same reason that the vacuum is trivial; so, the latter need not be assumed.
Discussing vacuum triviality  is useful chiefly in interpreting the FF results.

\section{Standard derivations of the Unruh effect\label{standard derivations}}
\subsection{Analyticity Argument\label{Analyticity Method}}
Here, we follow a standard IF approach to the Unruh effect \cite{Carroll:2004st} and simultaneously present the analogue in FF dynamics.
Herein and in the subsequent sections, we use {\color{red}red fonts} for IF-specific formulae, which are taken directly from Ref.\,\cite{Carroll:2004st}, and {\color{blue}blue fonts} for their FF equivalents.
Since the Unruh effect is usually discussed in the ``relativity convention'' metric, \textit{viz}.\ $(-,+,+,+)$, while the FF is generally discussed in the ``particle physics convention'' $(+,-,-,-)$, then to preserve the familiar formulae,
we use {\color{red}$(-,+,+,+)$} in the IF case and  {\color{blue}$(+,-,-,-)$} in FF.

Following Ref.\,\cite{Carroll:2004st}, we consider massless scalar particles in (1+1)D spacetime for simplicity.
The spacelike line element is thus:
\begin{equation}
\label{eq:2dmetric}
\begin{array}{llll}
{\color{red} ds^2=-dt^2+dz^2} & \rm{(IF)} ;
& \quad {\color{blue}ds^2=d\tau dx^-} &   \rm{(FF)}.
\end{array}
\end{equation}
A uniformly accelerated observer follows a trajectory:
\begin{equation}
\label{eq:Rindler-IF-FF-relations}
\begin{array}{llll}
\displaystyle
{\color{red} t(\rho)  =  \frac{1}{\alpha} \sinh(\alpha \rho)}  {\rm (IF)};~ 
\displaystyle 
{\color{blue} \tau(\rho)   \equiv   t(\rho)  +   z(\rho)   =  \frac{1}{\alpha}e^{\alpha \rho}}
 {\rm (FF)}, \\
\vspace{-2mm} \\ 
\displaystyle{\color{red} z(\rho)   =   \frac{1}{\alpha}  \cosh(\alpha \rho)}  {\rm (IF)}; 
\displaystyle{\color{blue} x^-(\rho)   \equiv   t(\rho)  -  z(\rho) 
 =  \frac{1}{\alpha}e^{-\alpha \rho}} 
{\rm (FF)},
\end{array}
\end{equation}
where $\alpha$ is the acceleration and $\rho$ a parameter. 
The trajectory describes a {\color{red} hyperboloid} in the IF and a {\color{blue} hyperbola} in the FF:
\begin{equation}
\label{eq:trajectories}
\begin{array}{llll}
\displaystyle
{\color{red} t^2(\rho)  =  z^2(\rho) - \alpha^{-2}} & {\rm (IF)}; &
\displaystyle
{\color{blue}\tau(\rho)  =  \frac{1}{\alpha^2 x^-(\rho)}} & {\rm (FF)}.
\end{array}
\end{equation}

Working with these kinematics, Rindler coordinates $\eta$, $\xi$ can be defined, along with the Rindler frame. 
Their relation to the Minkowski (\textit{i.e}., inertial) frame is:
\begin{equation}
\label{eq:Rindler frame I}
\begin{array}{llll}
\displaystyle
{\color{red} t  =  \frac{1}{a}e^{a\xi}\sinh(a \eta)} & {\rm(IF)}; 
&
\displaystyle
{\color{blue} \tau  =  \frac{1}{a}e^{a(\eta+\xi)}}& {\rm(FF)},\\
\vspace{-2mm} \\ 
\displaystyle
{\color{red} z  =  \frac{1}{a}e^{a\xi}\cosh(a \eta)} & {\rm(IF)}; 
&
\displaystyle
{\color{blue} x^-  =  \frac{1}{a}e^{a(-\eta+\xi)}}& {\rm(FF)},
\end{array}
\end{equation}
with the metric
$ds^2=e^{2a\xi}(-d\eta^2+d\xi^2).$
Thus, $\eta$ can be identified as the Rindler proper time and $\xi$ as the Rindler space coordinate.

 \begin{figure}[t]
 \center
\includegraphics[width=0.45\textwidth]{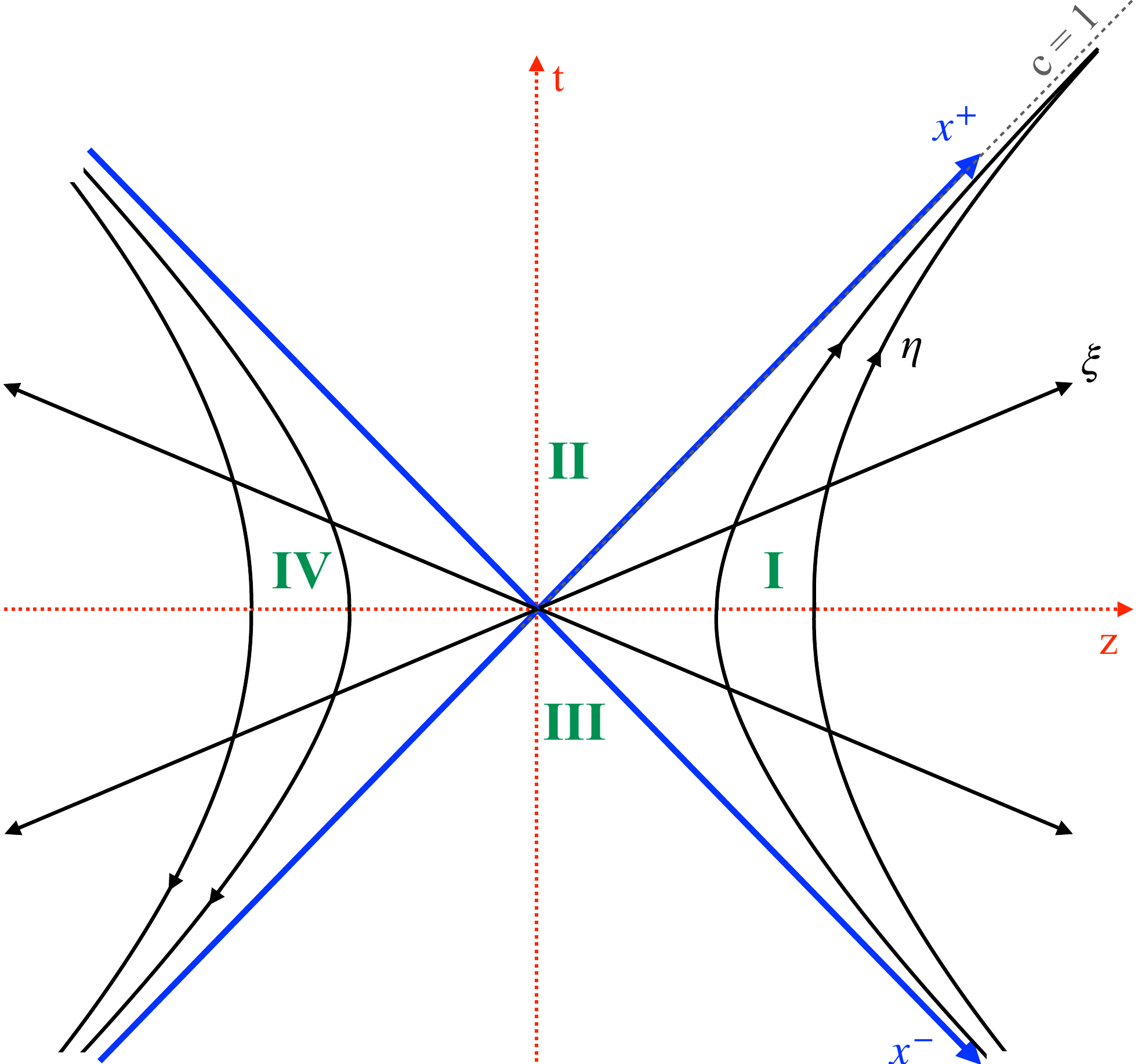}
\caption{Coordinate frames: Rindler (black), Minkowski IF (red), and FF (blue). 
As in Fig.~\ref{fig:IF_and_FF}, 
we use the $(-,+,+,+)$ metric for both IF and FF.
\label{fig:Rindler frame}}
\end{figure}

Equations~(\ref{eq:Rindler frame I}) provide a coordinate system only on the right spacelike quadrant of the Minkowski space delimited by $|t|=z, z\geq0$ and denoted quadrant {\bf I} in Fig.~\ref{fig:Rindler frame}.
Coordinates for the left spacelike quadrant ($|t|=z, z\leq0 $, quadrant {\bf IV}), are obtained by flipping Minkowski coordinate signs. This is because the Rindler space coordinate $\xi$ is a ray originating at $z=0$ rather than a line spanning $[-\infty,+\infty]$.
Therefore, in quadrant {\bf IV}, $\xi$ points left and the Rindler time is reversed: $\eta$ flows downward.
The reasoning applies to IF and equally to FF
since the FF Killing field is
\begin{equation}
{\color{blue}
\partial_\eta = \frac{\partial \tau}{\partial \eta} \partial_\tau + \frac{\partial x^-}{\partial \eta} \partial_{x^-} = a( \tau \partial_\tau - x^- \partial_{x^-})}\,,
\label{eq:Killing_field_FF}
\end{equation}
with components {\color{blue}$(2a\tau,-2ax^-)$}. Thus, $\partial_\eta$ is orthogonal to the  $\tau=0$ hypersurface, as in the IF case and, like the IF, $\partial_\eta$ defines positive- and negative-modes for a Fock basis.
Consequently, in quadrant {\bf IV},
\begin{equation}
\label{eq:Rindler frame IV}
\begin{array}{llll}
\displaystyle
{\color{red}  t  =  -  \frac{1}{a}e^{a\xi} \sinh(a \eta)} & {\rm(IF)}; 
&
\displaystyle
{\color{blue} \tau  = -  \frac{1}{a}e^{a(\eta+\xi)}}& {\rm(FF)},\\
\vspace{-2mm} \\ 
\displaystyle
{\color{red}   z  = -  \frac{1}{a}e^{a\xi} \cosh(a \eta)} & {\rm(IF)}; 
&
\displaystyle
{\color{blue}   x^-  = -  \frac{1}{a}e^{a(-\eta+\xi)}}& {\rm(FF)}.
\end{array}
\end{equation}

In Rindler coordinates, the Klein-Gordon equation for massless free fields is:
\begin{eqnarray}
\label{eq:Klein-Gordon Rindler}
\Box \phi = \frac{1}{a}e^{-2a\xi}(-\partial^2_\eta +\partial^2_\xi) \phi = 0.
\end{eqnarray}
(The extension to massive fields is straightforward and without qualitative import.)
Solutions are given by planes waves: $g_p=(4\pi \omega)^{-\sfrac{1}{2}} {\rm e}^{-i \omega \eta + i p \xi}$, with $\omega = |p|$.
In Minkowski frames, the Klein-Gordon equation is:
\begin{equation}
\label{eq:Klein-Gordon Minkowski}
\begin{array}{llll}
  {\color{red} \Box \phi = (-\partial^2_t +\partial^2_z) \phi = 0} & {\rm (IF)}; 
{\color{blue}\Box \phi = (\partial^+ \partial^-)  \phi = 0} & {\rm (FF)}.
\end{array}
\end{equation}
Here, 
{\color{blue} $\displaystyle\partial^+ \equiv 2\sfrac{\partial}{\partial x^-}$}, 
{\color{blue} $\displaystyle\partial^- \equiv 2\sfrac{\partial}{\partial \tau}$}. 
They are the operators associated with FF momentum, $p^+ = \omega + p$, and FF energy, $p^- = \omega - p$, respectively.
The plane wave solutions of Eqs.~(\ref{eq:Klein-Gordon Minkowski}) are:
\begin{equation}
\label{eq: Minkowski modes}
\begin{array}{llll}
\displaystyle
  {\color{red} f_p \propto  {\rm e}^{-i \omega t + i p z}}
\,,\; {\color{red} f_p^\ast} &{\rm (IF)}; 
\displaystyle
{\color{blue} f_{p^+} \propto  {\rm e}^{-\sfrac{i}{2}(p^+x^-)}}\,,\;
{\color{blue}f_{p^+}^\ast} & {\rm (FF)}. 
\end{array}
\end{equation}
There is  no {\color{blue} ${\rm e}^{-\sfrac{i}{2}(p^-\tau)}$} component in the FF wave function because massless particles are also zero-energy in FF dynamics: 
{\color{blue} $p^-\equiv \omega-p=0$.} 

Since in quadrant {\bf IV}, future Minkowski time (IF or FF) flows in the ``$-\eta$'' direction -- see Fig.\,\ref{fig:Rindler frame}, two sets of positive frequency Rindler modes must be introduced: one for quadrant {\bf I} and the other for quadrant {\bf IV},
\begin{subequations}
\label{eq:Rindler modes}
\begin{eqnarray}
g_p^{({\rm I})}=
\begin{cases}
      \displaystyle\frac{1}{\sqrt{4\pi \omega}}{\rm e}^{-i\omega \eta + ip\xi} & \rm I\\
      0  & \rm IV \\
    \end{cases}  \,,   \\
g_p^{({\rm IV})}=
\begin{cases}
      0  & \rm I \\
      \displaystyle\frac{1}{\sqrt{4\pi \omega}}{\rm e}^{+i\omega \eta + ip\xi} & \rm IV\\
          \end{cases}    \,,  
          \label{eq:Rindler modesIV}
\end{eqnarray}
\end{subequations}
so that $\omega \ge 0$ in both quadrants. 
The modes $g_p^{({\rm I})}$ and $g_p^{({\rm IV})}$ can be analytically continued outside their respective quadrant, which enables $\phi$ to be expressed in terms of four Rindler creation/annihilation operators:
\begin{eqnarray}
\label{eq:Rindler field}
\displaystyle
   \phi   =   \int   dp \big( \hat b_p^{({\rm I})}g_p^{({\rm I})}  +  \hat b_p^{({\rm I})\dag}g_p^{({\rm I})*}  +  \hat b_p^{({\rm IV})}g_p^{({\rm IV})}  +  \hat b_p^{({\rm IV})\dag}g_p^{({\rm IV})*} \big)  .
\end{eqnarray}

In contrast, Minkowski space not being partitioned, two creation/annihilation operators are sufficient:
\begin{equation}
\label{eq:Minkowski field}
\begin{array}{llll}
{\color{red} \phi = \int dp \big( \hat a_p f_p + \hat a_p^\dag f_p^*  \big)}  {\rm (IF)}; ~~~
{\color{blue} \phi = \int dp^+ \big( \hat a_{p^+} f_{p^+} + \hat a_{p^+}^\dag f_{p^+}^*  \big)}  {\rm (FF)} \,.
\end{array}
\end{equation}

The annihilation operators of Eqs.~(\ref{eq:Rindler field}( and (\ref{eq:Minkowski field}) define the Rindler and Minkowski vacua, respectively:
\begin{equation}
\label{eq:vacua}
\begin{array}{llllll}
\hat b^{({\rm I})}_p ~\ket{0_R}= \hat b^{({\rm IV})}_p \ket{0_R}= 0  {\rm (Rindler)}; ~~~
{\color{red} \hat a_p \hspace{2mm} \ket{0_M} =0 }  {\rm (IF)}; ~~~
{\color{blue} \hat a_{p^+} \ket{0_M} =0}  {\rm (FF)}.
\end{array}
\end{equation}

Exposing an Unruh effect hinges on showing that $\hat b^{({\rm I,IV})}_p$ cannot be expressed with $\hat a$ only but must also include $\hat a^\dag$. Then, a Rindler observer will detect particles from  the Minkowski vacuum,  $\hat b^{({\rm I,IV})}_p \ket{0_M} \neq 0$.

The expressions of $\hat b^{({\rm I,IV})}_p$ in terms of $\hat a$ and $\hat a^\dag$ can be obtained by finding Rindler modes $h^{({\rm I,IV})}_p$ that have the same vacuum as the Minkowski modes  $ f^{({\rm I,IV})}$ \cite{Unruh:1983ms}.
Using Eqs.~(\ref{eq:Rindler frame I}) and (\ref{eq:Rindler frame IV}), the relationship between the phases of the Rindler and Minkowski modes in the two spacetime quadrants is (recall the different metric signatures for IF and FF):
\begin{equation}
\label{eq:behavior-equiv}
\begin{array}{ll}
  {\rm e}^{-a(\eta-\xi)} 
&   = \left\{
\begin{array}{lllll}
{\color{red}-a(t-z)} & {\rm (IF)}; & 
\quad 
{\color{blue}   \phantom{-}a x^-} & {\rm (FF)} & \quad {\rm I} \\[1ex]
{\color{red} \phantom{-}a(t-z)} & {\rm (IF)}; & 
\quad 
{\color{blue}   - a x^-} & {\rm (FF)} & \quad {\rm IV}
\end{array}
\right. \\[1ex]
  {\rm e}^{a(\eta+\xi)}
&   = \left\{
\begin{array}{lllll}
{\color{red}\phantom{-}a(t+z)} & {\rm (IF)}; & 
\quad 
{\color{blue}   \phantom{-}a \tau}\phantom{^-} & {\rm (FF)} & \quad {\rm I} \\[1ex]
{\color{red} -a(t+z)} & {\rm (IF)}; & 
\quad 
{\color{blue}   - a \tau} & {\rm (FF)} & \quad {\rm IV}
\end{array}
\right. 
\end{array}\,
\end{equation}
Above, for the sake of generality, we wrote {\color{blue}$a\tau$}, but for a (1+1)D massless scalar field, contributions involving {\color{blue}$\tau$} vanish because their phase is {\color{blue}$p^-\tau$}, with {\color{blue}$p^-=0$}.
Recall also that $a$ is the acceleration and should not be confused with the operators $\hat a$, $\hat a^\dagger$.

We can now identify which Rindler modes match the spacetime dependence of the Minkowski modes {\color{red} $f_p$} or  {\color{blue} $f_{p^+}$}.
Using Eqs.~(\ref{eq:Rindler modes}) and (\ref{eq:behavior-equiv}) and the Rindler dispersion relation, $\omega=p$, the Rindler mode in quadrant {\bf I} that has the same phase as {\color{red} $f_p$} or {\color{blue} $f_{p^+}$} is:
\begin{equation}
\label{eq:9.152Carroll}
\begin{array}{llll}
{\color{red}\sqrt{4\pi\omega}g_p^{({\rm I})}=a^{i\omega/a}(-t+z)^{i\omega/a}}
{\rm (IF)}; ~~~
{\color{blue} \sqrt{4\pi\omega}g_p^{({\rm I})}=(ax^-)^{ip^+/2a}}  {\rm (FF)}.
\end{array}
\end{equation}

The available Rindler mode in quadrant IV, \eqref{eq:Rindler modesIV}, does not have the same phase as in \eqref{eq:9.152Carroll}.
To match the phase, we must consider the complex conjugate of $g_p^{({\rm IV})}$ with negative momentum $k := -p<0$:
\begin{equation}
\label{eq:9.154Carroll}
\begin{array}{llll}
{\color{red}\sqrt{4\pi\omega}g_{-p}^{({\rm IV})*}=a^{i\omega/a}e^{\pi \omega /a}(-t+z)^{i\omega/a}}
 {\rm (IF)}; ~~~
{\color{blue} \sqrt{4\pi\omega}g_{-p}^{({\rm IV})*}=e^{\pi p^+ /2a}(ax^-)^{ip^+/2a}}  {\rm (FF)}.
\end{array}
\end{equation}

The Rindler dispersion relation, $\omega=|k|$, and the definition, {\color{blue}$p^+ \equiv \omega + k $}, impose {\color{blue}$p^+ =0$} on the domain $k=-p \leq 0$.
Thus, in the FF case, there is no available Rindler mode in quadrant {\bf IV}.
\eqref{eq:9.154Carroll} then becomes:
\begin{equation}
\label{eq:9.154Carroll-2}
\begin{array}{llll}
{\color{red}\sqrt{4\pi\omega}g_{-p}^{({\rm IV})*}=a^{i\omega/a}e^{\pi \omega /a}(-t+z)^{i\omega/a}}
 {\rm (IF)} ; ~~~
{\color{blue} \sqrt{4\pi\omega}g_{-p}^{({\rm IV})*}
= {\rm constant} = 0 }
 {\rm (FF)}.
\end{array}
\end{equation}
The rightmost result follows because $g_{-p}$ must be a plane wave, by definition of the Fourier decomposition of $\phi$, \eqref{eq:Rindler field}, and a nonzero value for the constant would violate this constraint.

With Eqs.~(\ref{eq:9.152Carroll}) and (\ref{eq:9.154Carroll-2}), the Rindler modes can now be expressed in terms of the Minkowski modes:
\begin{equation}
\label{eq:9.155Carroll}
\begin{array}{llll}
 {\color{red}\sqrt{\pi\omega} \big( g_p^{({\rm I})}+e^{-\pi\omega/a}g_{-p}^{({\rm IV})*} \big) = a^{i \omega/a}\big(-t+z \big)^{i \omega/a}} 
 {\rm (IF)}; ~~~
 {\color{blue}
\sqrt{4\pi\omega}g_p^{({\rm I})}=(ax^-)^{ip^+/2a}}
 {\rm (FF)}.
\end{array}
\end{equation}
Plainly, only the Rindler mode $g^{({\rm I})}_p$ matches the phase of the Minkowski FF modes {\color{blue}$f_{p^+}$} -- see \eqref{eq: Minkowski modes}, and not $g^{({\rm IV})*}_{-p}$; the Bogolyubov matrix is diagonal. Hence, there is no Unruh effect in the FF framework.
Notwithstanding this, concluding details are included below.

The IF combination in \eqref{eq:9.155Carroll} is not yet properly normalized. The normalized IF mode is (the FF mode remains unchanged):
\begin{equation}
\label{eq:9.156Carroll}
\begin{array}{llll}
\displaystyle
  {\color{red}h_p^{({\rm I})}=\frac{1}{\sqrt{2 \sinh(\sfrac{\pi \omega}{a})}}  \big( e^{\pi\omega/2a} g_p^{({\rm I})} + e^{-\pi\omega/2a}g_{-p}^{({\rm IV})*} \big)}
  {\rm (IF)}; 
  \displaystyle
  {\color{blue}~ g_p^{({\rm I})} =
 \frac{1}{\sqrt{4\pi\omega}} \big( ax^- \big)^{i p^+/2a}}
   {\rm (FF)}.
\end{array}
\end{equation}

The normalized Rindler modes, $h_p$, have corresponding creation/annihilation operators $\hat c_p$, which, by the construction above, act on the Minkowski vacuum as $\hat c_p \ket{0_M}= 0$ and $\hat c^\dagger_p \ket{0_M}= \ket{p}$.
Using \eqref{eq:9.156Carroll} and analogous reasoning for $\hat b_p^{({\rm IV})}$, one obtains the relationship between $\hat c_p$, $\hat c^\dagger_p$ and the original Rindler operators:
\begin{equation}
\label{eq:9.160Carroll}
\begin{array}{llll}
\displaystyle
{\color{red}   \hat b^{({\rm I})}_p = \frac{1}{\sqrt{2 \sinh(\sfrac{\pi \omega}{a})}}  \big( {\rm e}^{\pi\omega/2a} \hat c_p^{({\rm I})} + {\rm e}^{-\pi\omega/2a} \hat c_{-p}^{({\rm IV})\dagger} \big)} & {\rm (IF)}; ~~~
{\color{blue}   \hat b^{({\rm I})}_p = \hat a_{p^+}} & {\rm (FF)}, \\
\displaystyle
{\color{red}   \hat b^{({\rm IV})}_p  =  \frac{1}{\sqrt{2 \sinh(\sfrac{\pi \omega}{a})}}  \big( {\rm e}^{\pi\omega/2a} \hat c_p^{({\rm IV})} + {\rm e}^{-\pi\omega/2a} \hat c_{-p}^{({\rm I})\dagger} \big)} & {\rm (IF)}; ~~~
\displaystyle
{\color{blue}   \hat b^{({\rm IV})}_p = \hat a_{p^-}} & {\rm (FF)}.
\end{array}
\end{equation}

Acting now with, \textit{e.g}., $\hat b^{({\rm I})}_p$ on the Minkowski vacuum, one finds:
\begin{equation}
\label{eq: Unruh effect}
\begin{array}{llll}
\displaystyle
{\color{red} \hat b^{({\rm I})}_p     \ket{0_M}     =     \frac{1}{\sqrt{2 \sinh(\sfrac{\pi \omega}{a})}}     \big( {\rm e}^{-\pi\omega/2a}     \ket{-p}     \big)}
 {\rm (IF)}; ~~~
\displaystyle
{\color{blue} \hat b^{({\rm I})}_p     \ket{0_M}     =     \hat a_{p^+} \ket{0_M} =0} 
{\rm (FF)}.
\end{array}
\end{equation}
Evidently, in a Rindler frame, following this reasoning, one can observe particles from the IF Minkowski vacuum but not from the (trivial) FF vacuum.

The crucial point is the transition between Eqs.~(\ref{eq:9.152Carroll}) and (\ref{eq:9.155Carroll}): modes with negative momentum, $k=-p$, are introduced.
This is allowed in the IF: it simply means waves moving toward ``$-z$'' (for quadrant I); and they can be detected by an observer in that quadrant.
In the FF, on the other hand, {\color{blue}$p^+ > 0$} always;
waves with {\color{blue}$p^+ \leq 0$} would propagate backward in IF time; thus violating causality.
Therefore, they do not exist; so, neither do Unruh particles.
In other words, because, formally, $f_{-p^+} = f^\ast_{p^+}$, one must set  $f_{-p^+} = 0$ or one would double count in the Fourier expansion, \eqref{eq:Minkowski field}.
(Note that $f^\ast_{p^+}$ cannot be invoked for \eqref{eq:9.154Carroll} since the procedure is to match Rindler modes to $f_{p^+}$, not to $f^\ast_{p^+}$.)

The demonstration provided in this Appendix is formal. 
To apprehend the difference between the IF and FF demonstrations, it is helpful to interpret the central result, \eqref{eq:9.156Carroll}. 
Let us first consider the usual mode expansion of a field operator, \eqref{eq:Minkowski field}].
Modes $f_p$ and $f_p^*$ correspond to particles and antiparticles since $\hat a_p$ removes a quantum of the field and $\hat a_p^\dag$ adds one. 
Let us choose to call $f_p$ the antiparticle mode and $f_p^*$ the particle mode 
\eqref{eq:9.156Carroll} for IF, \textit{viz}.\ 
$ {\color{red}h_p^{({\rm I})} \propto  \big( e^{\pi\omega/2a} g_p^{({\rm I})} + e^{-\pi\omega/2a}g_{-p}^{({\rm IV})*} \big)}$, 
means that for an individual mode $h_p$, particles of momentum $-p$ and antiparticles of momentum $p$ co-exist, as in the disconnected pair-creation diagrams of vacuum loops.  
This is the standard interpretation of the Unruh effect: the IF vacuum structure, made of disconnected loops of particles/antiparticles of opposite momenta, becomes observable. 
In the FF case, the term {\color{blue} $e^{-\pi\omega/2a}g_{-p}^{({\rm IV})*}=(ax^-)^{ip^+/2a}$} does not contribute because its power, {\color{blue} $p^+:= \omega+(-p)=0$}, and a constant mode normalizes to 0. 

The absence of $g_{-p}^{({\rm IV})*}$ does not permit the simultaneous creation of a particle and an antiparticle with opposite momenta, {\it i.e.}, of disconnected vacuum loops, as is well-known for FF.

\subsection{Direct derivation of the Bogolyubov transformation\label{Direct Bogolyubov transform}}
As discussed in Appendix~\ref{Analyticity Method}, the Bogolyubov transformation was derived following the original analyticity argument \cite{Unruh:1983ms}. 
Here we discuss the direct derivation -- see, \textit{e.g}., Ref.~\cite{Crispino:2007eb}. 
The steps need not be fully detailed because the argument precluding the
Unruh effect in FF is the same as in Appendix~\ref{Analyticity Method}.

The Bogolyubov transformation is 
\begin{equation}
\label{eq:Crispino-Eq. 2.53}
\begin{array}{llllll}
\displaystyle
g_p^{({\rm I})}  = 
\displaystyle\int_0^\infty \frac{d\omega}{\sqrt{4\pi \omega}} \big( \hat \alpha_p^{({\rm I})} f_p + \hat \beta_p^{({\rm I})} f_p^* \big)\,, ~~~
g_p^{({\rm IV})}  =  
\displaystyle\int_0^\infty \frac{d\omega}{\sqrt{4\pi \omega}} \big( \hat \alpha_p^{({\rm IV})} f_p + \hat \beta_p^{({\rm IV})} f_p^* \big)\,,
\end{array}
\end{equation}
with
$\hat \beta_p^{({\rm IV})} = -{\rm e}^{-\pi\omega/a}  \hat \alpha_p^{({\rm I})\ast}$, 
$\hat \beta_p^{({\rm I})} = -{\rm e}^{-\pi\omega/a}  \hat \alpha_p^{({\rm IV})\ast}$.

For IF, these coefficients are non-zero; hence, the Unruh effect is predicted. 
In contrast, for FF, $\hat \beta_p^{({\rm IV})}$$=\hat \alpha_p^{({\rm IV})*}=0$ because no FF modes are permitted in quadrant {\bf IV}.  
Hence, $\hat \beta_p^{({\rm IV})} = \hat \beta_p^{({\rm I})}=0$ and no Unruh effect is perceived.

\section{The Unruh effect in thermal quantum field theory\label{Unruh and thermal QFT} }

Before examining the Unruh effect using thermal QFT, it is worth recalling the correspondence between Rindler time translations and Minkowski boosts, and the implications for Rindler Hamiltonian definition in FF dynamics.

\subsection{Relation between Rindler time translations and boosts: implications for the Hamiltonian\label{boost-H argument}}
For IF and FF, a Rindler time translation corresponds to a Minkowski boost.
This is readily seen using Eqs.~(\ref{eq:Rindler frame I}) and making a Rindler time translation $\eta \to \eta +\Delta$, which leads to:
\begin{equation}
\label{eq:R-dt=M-boosts-1}
\begin{array}{llll}
\displaystyle
{\color{red}t^\prime = \frac{1}{a}{\rm e}^{a\xi}\sinh(a \eta+a\Delta)}, \\
\vspace{-2mm} \\ 
{\color{red}~~~=\frac{1}{a}{\rm e}^{a\xi}\big(\sinh(a \eta)\cosh(a\Delta) + \cosh(a \eta)\sinh(a\Delta) \big) }
\quad \rm{(IF)}; \\
\vspace{-2mm} \\ 
\displaystyle
{\color{blue} {\tau}^\prime=\frac{1}{a}e^{a(\eta+\Delta+\xi)}}
 \qquad \rm{(FF)}, \\
\vspace{-2mm} \\ 
\displaystyle
{\color{red} z^\prime=\frac{1}{a}{\rm e}^{a\xi} \cosh(a \eta+a\Delta) }, \\
\vspace{-2mm} \\ 
{\color{red}~~~= \frac{1}{a}{\rm e}^{a\xi}\big(\cosh(a \eta)\cosh(a\Delta) + \sinh(a \eta)\sinh(a\Delta) \big)}  \quad \rm{(IF)}; \\
\vspace{-2mm} \\ 
\displaystyle
{\color{blue} {x^-}'= \frac{1}{a}{\rm e}^{a(-\eta-\Delta+\xi)} }  \quad\rm{(FF)}.
\end{array}
\end{equation}
Hence, 
\begin{equation}
\label{eq:R-dt=M-boosts-2}
\begin{array}{llll}
\displaystyle
{\color{red} t^\prime = t\cosh(a\Delta) + x\sinh(a\Delta)} & \mathrm{(IF)}; 
\displaystyle
{\color{blue} \tau^\prime = \mathrm{e}^{a\Delta}\tau} & \mathrm{(FF)},\\
\displaystyle
{\color{red} z^\prime = z\cosh(a\Delta) + t\sinh(a\Delta)} & \mathrm{(IF)}; 
\displaystyle
{\color{blue} {x^-}^\prime = e^{-a\Delta}x^-} & \mathrm{(FF)},
\end{array}
\end{equation}
which are the usual boost formulae with rapidity $a\Delta$. 

Now, since FF boosts are kinematical, the above correspondence makes, in FF dynamics, the Rindler time translation a kinematical operator. Therefore, it cannot serve as the Hamiltonian.
In Euclidean space, where thermal QFT is used to study the Unruh effect, the dilation operator defines the FF Hamiltonian, as explained in the next section.

\subsection{Thermal QFT}
For a scalar field, the thermal QFT derivation of the IF Unruh effect \cite{Bisognano:1975ih} utilizes the periodicity in Euclidean time (\textit{viz}.\ imaginary time $\tilde{t}: = i t$) of a QFT at finite temperature $T=1/\beta$, $0\leq \tilde t \leq  \beta$.

Consider the two-point Schwinger function of field $\phi(\tilde{t},z)$:
\begin{eqnarray}
\label{eq:G's function-1}
G_\beta(\tilde{t},z) = \frac{1}{Z}{\rm Tr}~{\rm e}^{- \beta H} \mathcal T [ \phi(\tilde{t},z) \phi(0,0)],
\end{eqnarray}
with $Z$ the partition function and $\mathcal T$ the imaginary-time ordering operator.
In the Boltzmann factor ${\rm e}^{-\beta H}$, $H$ is the operator providing energy eigenvalues, {\it not necessarily the time-evolution operator}, although
it is both in IF dynamics.
This implies  {\color{red}${\rm e}^{-\beta H} \phi(\tilde t,z) {\rm e}^{ \beta H} = \phi(\tilde t - \beta,z)$}, which, with the cyclicity property of a trace and \eqref{eq:G's function-1}, yields -- see, \textit{e.g}., Ref.\,\cite[Eq.\,(2.86)]{Kapusta:1989}:
\begin{eqnarray}
\label{eq:G's function-2}
 {\color{red}G_\beta(\tilde{t},z)=G_\beta(\tilde{t}-\beta,z).}
\end{eqnarray}

Using Eqs.~(\ref{eq:R-dt=M-boosts-2}), a shift in imaginary time
$\tilde t \to \tilde t + i\Delta $, $i\Delta = 2\pi/a$, is seen to yield
{\color{red} $t^\prime=t$ and $z^\prime=z$}
or
{\color{blue} ${\tau}^\prime =\tau$ and ${x^-}^\prime =x^-$}.
Namely, after a Wick rotation, the Rindler metric makes a succession of boosts that correspond to a single rotation in imaginary time; and the shift is just a change in temperature.
On the other hand, in FF dynamics, as shown in Section~\ref{boost-H argument}, the Rindler Hamiltonian is not the time translation
operator; so,
{\color{blue}${\rm e}^{-\beta H} \phi(0,0) {\rm e}^{\beta H}\neq \phi(\beta,0)$} and there is no cyclic relation in FF.

Evidently, only in IF does the formal equivalence between $\beta$ and periodicity yield an Unruh temperature {\color{red} $\beta^{-1}$$=$$T$$=$$a/2\pi$}.

The Rindler frame identifies with Dirac's Point-Form (PF, see Section~\ref{Dirac's Forms of Dynamics}) 
after switching time and space, which is nugatory in Euclidean space because the axes are indistinguishable.
The PF Hamiltonian being the dilation operator \cite{Fubini:1972mf}, as evinced by the continuous dilation of the axes attached to the red wordline in Fig.\,\ref{fig:IF_and_FF}, 
right panel, one may choose it as the FF Rindler Hamiltonian.
One then recovers \eqref{eq:G's function-2}, but the Rindler frame is no longer equivalent to a cyclic frame under dilation factors $d_{1,2}$:
${\tau}'=\frac{1}{a}e^{a(d_1 \eta+ d_2\xi)} \neq \tau$ except for the trivial case $d_1=1=d_2$.
Consequently, the thermal QFT approach reveals no Unruh effect in FF dynamics.

\section{Dynamical derivation\label{Dynamical derivation}}
Consider the Lagrangian of a free 
scalar field $\varphi$ in (1+1)D:
\begin{equation}
\label{eq: 2D model Lagrangian-free}
\mathcal{L}=\frac{1}{2} \partial_\mu \varphi \partial^\mu \varphi - \frac{1}{2}m^2 \varphi^2,
\end{equation}
with $m$ the field mass.
Then change the perspective to a frame in which the field appears shifted,
\begin{equation}
\label{eq: field shift}
 \varphi \equiv \phi+ a/m^2.
\end{equation}
In this frame, after subtracting a (physically immaterial) constant, 
$a^2/[2m^2]$, 
$\mathcal{L}$ becomes:
\begin{equation}
\label{eq: 2D model Lagrangian}
\mathcal{L}=\frac{1}{2} \partial_\mu \phi \partial^\mu \phi - \frac{1}{2}m^2 \phi^2 +a \phi \,,
\end{equation}
where $a$ acts as an external source term and provides a linear potential, \textit{viz}.\ a constant  acceleration.
This model was studied in Refs.\,\cite{Heinzl:1998kz, Hornbostel:1991qj}, albeit not in the context of the Unruh effect.
To see that $a \phi$ in \eqref{eq: 2D model Lagrangian} produces a constant force, consider the classical mechanics equivalent, $\mathcal L =\frac{1}{2}m\dot z^2-\frac{1}{2}kz^2-az$, where $k$ is the harmonic oscillator constant.
The potential term $a z$ generates a constant force $a$ in the (1+1)D case \cite{note-pot.term}.

Suppose a detector sensitive to $\phi$ is attached to the new frame.
It may be pictured as being composed of fermions that couple to $\phi$.
The field waves $\partial_\mu \phi \partial^\mu \phi$ induce transitions in the detector, making it sensitive  to $\phi$, and the $a \phi$ term accelerates the detector.

To avoid infrared issues, suppose space is of finite extent, with length $L$ and circular boundary conditions.
Then decompose the two fields as follows (the $p^\mu_n$ spectrum is discrete because $L$ is finite):
\begin{subequations}
\label{eq:field exp. a,a+}
\begin{align}
\phi = \frac{1}{\sqrt{2L}} \sum_n \big( \hat A_{p_n} e^{ip^\mu_n x_\mu} + \hat A_{p_n}^\dag e^{-ip^\mu_n x_\mu}  \big),  ~~~
\varphi  = \frac{1}{\sqrt{2L}} \sum_n \big( \hat a_{p_n} e^{ip^\mu_n x_\mu} + \hat a_{p_n}^\dag e^{-ip^\mu_n x_\mu}  \big).
\end{align}
\end{subequations}

The relation between $(\hat A_{p_n},\hat A^\dag_{p_n})$ and $(\hat a_{p_n}$,$\hat a^\dag_{p_n})$ is obtained by expressing \eqref{eq: field shift} in terms of a translation operator, $U$:
\begin{equation}
\label{eq: field shift-2}
 \varphi  = U^\dag \phi U, \, \quad U = {\rm e}^{-ia{\mathsf P}/m^2}\,,
\end{equation}
where
$\mathsf P$ is the momentum integrated over a time $\Delta T$ characterizing the (Unruh) detector response. $\mathsf P$  is obtained from the momentum density, \textit{viz}.\ the canonical conjugate of $\varphi$:
\begin{equation}
\label{eq:momentum density}
{\mathsf p} = \frac{-i}{\sqrt{2L}} \sum_n \sqrt{\frac{\omega_n}{2}} \big( \hat a_{p_n} {\rm e}^{ip^\mu_n x_\mu} - \hat a_{p_n}^\dag {\rm e}^{-ip^\mu_n x_\mu}  \big).
\end{equation}

Focusing now on IF dynamics (red fonts):
\begin{equation}
\label{eq:averaged mom. IF}
   {\mathsf P} = \int_{\Delta T, \pm L}     d^2 x \, {\mathsf p}    =   
   {\color{red}  -i \sum_n \sqrt{\omega_nL} \Delta T  \big( \hat a_{p_n} - \hat a_{p_n}^\dag \big)}.
\end{equation}
(Factors of $\pi$ are omitted here as they are immaterial to the discussion.)
Equations~(\ref{eq: field shift-2}) and (\ref{eq:averaged mom. IF}) give
\begin{equation}
\label{eq:translation ops}
   {\color{red}   U = \exp{\displaystyle\frac{a \Delta T}{m^2} \sum_n \sqrt{\omega_nL}  \big( \hat a_{p_n} - \hat a_{p_n}^\dag \big)}},
\end{equation}
which becomes, upon using the Baker-Campbell-Hausdorff formula and the commutation properties of $\hat a_{p_n}$ and $\hat a^\dag_{p_n}$:
\begin{equation}
\label{eq:translation ops-2}
   {\color{red}  U    =     \prod_n
   \exp     \left[\frac{(a \Delta T)^2}{2m^4} \omega_nL\right]
    \exp     \left[ - \frac{a \Delta T\sqrt{\omega_nL}}{m^2}   \hat a_{p_n}\right]} .
\end{equation}

The Unruh effect is a statement that although the vacuum of an inertial field (here $\varphi$) is, by definition, without real particles, so that $\hat a_{p_n} \ket{0_M} \equiv 0$, an accelerated field (here $\phi$) is not:
$ {\color{red}  \hat A_{p_n} \ket{0_M} = U\hat a_{p_n} U^\dag \ket{0_M} \neq 0 }$.
This again suggests that the IF predicts an observable Minkowski vacuum.
However, the detector involved in the observation process should also be included in the analysis.
Such would require incorporation of fermion fields into the Lagrangian, \eqref{eq: 2D model Lagrangian-free}, and subsequent study of the bound states constituting the detector.
The discussion Section~\ref{Gedanken experiment} 
then indicates that the interaction of $a\phi$ with the detector, \textit{i.e.},\ the dynamical effect of the time-dependent boost, will cancel the Unruh effect.

For the FF, however, the momentum density is a simple space derivative: 
{\color{blue} ${\sf p} = \partial^+ \varphi \equiv 
\partial \varphi/\partial x^- $}.
Therefore:
\begin{equation}
\label{eq:averaged mom. FF}
  {\mathsf P} =  \int_{\Delta T, \pm L}
  {\color{blue} d\tau dx^- \frac{\partial \varphi}{\partial x^- } }
  = {\color{blue} \int_{\Delta T}d\tau \big[ \varphi \big]_{-L}^{+L}}
  = {\color{blue} 0 },
\end{equation}
owing to the periodic boundary conditions.
Thus, $  {\color{blue} U = \mathds{1} }$ and the trivial vacuum of $\varphi$ in Minkowski space $\hat a_{p_n} \ket{0_M}=0$ remains such for the accelerated field:
$ {\color{blue}  \hat A_{p_n} \ket{0_M} = \hat a_{p_n} \ket{0_M} = 0 }$.
There is no Unruh effect in FF.

One may interpret the IF and FF results as follows.\\
{\color{red} (\textit{i}) IF. 
${\mathsf p}$ is the time derivative of the field, ${\mathsf p}=
\partial \varphi/ \partial t$.
Thus, it gives the difference between $\varphi$ on two infinitesimally separated hypersurfaces, \textit{i.e}., its dynamical evolution.}\\
{\color{blue} (\textit{ii}) FF. ${\mathsf p}$ is a space derivative of the field, ${\mathsf p} = \partial \varphi/\partial x^- $.  It
operates on a single hypersurface  without dynamical evolution.} \\
Consequently, the conclusions of the previous sections are recovered: (\textit{i}) the IF treatment  {\it dynamically} affects a seemingly nontrivial vacuum, whereas (\textit{ii}) the FF treatment does not impact upon a trival vacuum.
These qualities trace back to the dynamical vs. kinematical character of IF \textit{cf}.\ FF boosts.

\bibliographystyle{elsarticle-num-names}
\bibliography{BiBUnruh.bib}

\begin{thebibliography}{999}
\bibitem[EP-()]{EP-TD}
It also provides an intuitive understanding of the thermal distribution. The
  greater distance scale set by larger loops allows for a larger curvature,
  therefore, by the equivalence principle, a larger acceleration and, in turn,
  a larger Unruh temperature. Since large loops are less probable than the
  smaller loops, so also are larger temperatures.
\bibitem[not()]{note-pot.term}
In (1+2)D or (1+3)D spacetimes, $a \propto t$ and $a \propto t^2$,
  respectively, are needed to obtain a constant force. This may be understood
  as the flux from source $a$ diluting in 2D or 3D space. To keep a constant
  force on the accelerating body moving away from the source, the latter must
  intensify with $t$.
\end{thebibliography}

\end{document}